\documentclass[aps,pra,reprint,notitlepage,superscriptaddress]{revtex4-2}

\usepackage{amsmath,amssymb,bm,bbm}
\usepackage{amsthm}
\usepackage{graphicx}
\usepackage{hyperref}
\usepackage{orcidlink}
\usepackage{bbm}
\usepackage[capitalize]{cleveref}
\usepackage[justification=justified,singlelinecheck=false]{caption}
\usepackage{subcaption}
\usepackage[section]{placeins}
\usepackage{ragged2e}

\newcommand{\Tr}{\operatorname{Tr}}
\newcommand{\Ree}{\operatorname{Re}}
\newcommand{\Imm}{\operatorname{Im}}

\newcommand{\spr}{\operatorname{spr}} 

\newcommand{\dd}{\mathrm{d}}
\newcommand{\ii}{\mathrm{i}}
\newcommand{\e}{\mathrm{e}}

\newcommand{\vect}{\operatorname{vec}}
\newcommand{\T}{\mathsf{T}}

\newcommand{\Sp}{\mathrm{Sp}}
\newcommand{\diag}{\mathrm{diag}}

\theoremstyle{plain}
\newtheorem{theorem}{Theorem}
\newtheorem{proposition}[theorem]{Proposition}

\theoremstyle{remark}
\newtheorem*{remark*}{Remark} 

\theoremstyle{definition}

\DeclareMathOperator{\adj}{adj}

\begin{document}

\title{Exceptional points in Gaussian channels: diffusion gauging and drift-governed spectrum}

\author{Frank Ernesto Quintela Rodr\'iguez\,\orcidlink{0000-0002-9475-2267}}
\email[Email: ]{frank.quintela@uam.es}
\affiliation{Condensed Matter Physics Center (IFIMAC), Universidad Aut\'onoma de Madrid, E-28049 Madrid, Spain}

\date{\today}

\begin{abstract}
McDonald and Clerk [Phys.\ Rev.\ Research 5, 033107 (2023)] showed that for linear open quantum systems the
Liouvillian spectrum is independent of the noise strength. We first make this noise-independence principle precise in
continuous time for multimode bosonic Gaussian Markov semigroups: for Hurwitz drift, a time-independent Gaussian
similarity fixed by the Lyapunov equation gauges away diffusion for all times, so eigenvalues and
non-diagonalizability are controlled entirely by the drift, while diffusion determines steady
states and the structure of eigenoperators. We then extend the same separation to discrete time for general stable
multimode bosonic Gaussian channels: for any stable Gaussian channel, we construct an explicit Gaussian similarity
transformation that gauges away diffusion at the level of the channel parametrization. We illustrate the method with a
single-mode squeezed-reservoir Lindbladian and with a non-Markovian family of single-mode Gaussian channels, where the
exceptional-point manifolds and the associated gauging covariances can be obtained analytically.
\end{abstract}

\maketitle

\section{Introduction}

Exceptional points (EPs)---parameter values at which a non-Hermitian operator becomes non-diagonalizable and eigenmodes
coalesce---provide a sharp organizing principle for non-unitary dynamics, with signatures ranging from mode coalescence to
polynomially dressed transients when Jordan blocks form \cite{Kato1995,Heiss2012,ElGanainy2018,Ashida2020}. In open quantum
physics the relevant non-Hermitian objects are superoperators: the Liouvillian generator $\mathcal L$ for continuous-time
evolution and, more generally, the quantum channel $\Psi$ describing a finite-time map (including stroboscopic and
discrete-time protocols). While Liouvillian EPs have been explored in a variety of settings \cite{Minganti2019,Chen2022PRL},
the channel viewpoint is both more general and increasingly natural experimentally, since finite-time maps are directly
accessible (e.g.\ by process tomography) even when a time-local generator description is not available.

A particularly active direction concerns non-Markovian quantum exceptional points (QEPs), where memory effects preclude a
Gorini--Kossakowski--Sudarshan--Lindblad (GKSL) semigroup description and EP physics must be formulated in terms of
finite-time dynamical maps. Recent work has made clear progress on this front, including experimental observation of
non-Markovian QEPs in a controlled platform \cite{ZhangHanWuNingYangZheng2025NonMarkovianQEP} and complementary theory
clarifying EP conditions and spectral signatures in non-Markovian settings \cite{LinKuoLambertMiranowiczNoriChen2025}. These
results emphasize that EPs can arise even when one deals only with a family of intermediate maps $\{\Psi_t\}$, whose
eigenvalues and Jordan structure may vary nontrivially with the finite time $t$.

From a broader viewpoint, such non-Markovian scenarios sit inside the general framework of \emph{quantum-channel} EPs: one
asks when a completely positive and trace-preserving (CPTP) map $\Psi$ (or a fixed-time map $\Psi_t$) becomes
non-diagonalizable as a superoperator. Despite its conceptual naturalness and direct experimental accessibility, a
systematic EP theory formulated \emph{directly} at the channel level is still relatively scarce. A notable recent step is
the channel-interpolation framework of Wong--Zeng--Li, which constructs CPTP families that exhibit EPs and connects them to
finite-time implementations \cite{WongZengLi2025}. These developments motivate a basic structural question: for
experimentally relevant channel families, which part of the dynamics controls EP \emph{locations} and \emph{order}, and
which part merely dresses the associated eigenoperators and steady states?

Continuous-variable (CV) platforms offer a particularly transparent arena. Bosonic modes---optical and microwave cavities,
propagating fields, collective spin-wave modes in atomic ensembles, and motional/vibrational modes in optomechanics and
trapped ions---are ubiquitous across quantum optics and central to modern quantum technologies
\cite{GardinerZoller,Wallsmilburn,Weedbrook2012,SerafiniBook}. In many such systems the reduced dynamics is well captured
by Gaussian operations: quadratic Hamiltonians together with dissipation linear in the canonical quadratures generate
Gaussian diffusions, and the associated finite-time maps are bosonic Gaussian channels
\cite{Weedbrook2012,CarusoGiovannettiHolevo2014,SerafiniBook}.
Beyond their dynamical role, Gaussian channels are workhorses of
CV quantum information, underpinning communication links, sensing protocols, state preparation and stabilization, and
modular architectures for optical and microwave platforms \cite{Weedbrook2012,SerafiniBook}.

In the Weyl (symmetrically ordered) characteristic-function representation, a Gaussian channel is specified by a triple
$(X,Y,\delta)$: a linear phase-space map $X$ (drift), a diffusion matrix $Y$ (noise), and a displacement $\delta$. This
drift--diffusion structure suggests a natural hypothesis for EP physics: non-diagonalizability should be governed by the
drift, while diffusion should affect only the structure of eigenoperators and fixed points. The main contribution of this
work is to make this separation precise by constructing explicit Gaussian similarity transformations that \emph{gauge away
diffusion} from the eigenvalue problem, in both continuous and discrete time. In this way, EP searches reduce to
finite-dimensional linear algebra---defect sets of the drift matrices $A$ (generators) or $X$ (channels)---while diffusion
is relegated to solving Lyapunov/Stein equations that determine the dressing covariance governing eigenoperators and steady
states.

Our first main result concerns \emph{continuous-time} Gaussian Markov dynamics. For a Gaussian diffusion generator with
Hurwitz drift matrix $A$, the unique solution $S$ of the Lyapunov equation defines a time-independent Gaussian smoothing
map $\mathcal V_S$ that gauges diffusion out of the Ornstein--Uhlenbeck generator by similarity, and equivalently out of
the associated semigroup \emph{uniformly for all times}, yielding a diffusion-free representative
$\widetilde\Psi_t\equiv(\e^{At},0,\delta_t)$. This generator-level theorem ties together semigroup structure,
complete-positivity constraints, and the $(X,Y,\delta)$ parametrization in a unified phase-space language. As a direct
consequence, Liouvillian EP manifolds are drift-controlled (they coincide with the defect set of $A$), while diffusion
enters only through the dressing covariance $S$ that shapes eigenoperators and the steady state.

Our second main result is a \emph{discrete-time} theorem for a single stable Gaussian channel $\Psi\equiv(X,Y,\delta)$ with
\textit{spectral radius} $\spr(X)<1$.
We construct a Gaussian smoothing map $\mathcal V_S$ such that
$\widetilde\Psi=\mathcal V_S^{-1}\circ\Psi\circ\mathcal V_S$ has parameters $(X,0,\delta)$, where $S$ is uniquely fixed by
a Stein equation determined by $(X,Y)$. Since similarity preserves eigenvalues and Jordan structure, this shows that EP
locations and their order are governed by defectiveness of $X$, while diffusion reshapes eigenoperators and determines the
fixed point through the gauging covariance $S$. Although motivated by the continuous-time construction, this discrete-time
statement does not assume that $\Psi$ is embeddable in a Markov semigroup and is therefore logically independent of the
generator-level theorem.

This manuscript is organized as follows. In~\Cref{sec:gaussian_channels} we introduce bosonic Gaussian channels and recall
their characterization by $(X,Y,\delta)$ in the characteristic-function and moment representations.
We then prove our
noise-gauging results: a continuous-time theorem establishing Lyapunov gauging for Gaussian diffusion generators (and the
associated semigroups), and a discrete-time theorem showing that, for stable channels, the Stein solution produces a
Gaussian similarity transformation that removes diffusion from the eigenvalue problem while preserving eigenvalues and
Jordan structure.

In~\Cref{sec:example_single_mode_nm} we apply these results to single-mode settings. We first analyze a Markovian
squeezed-reservoir Lindbladian, where the drift exhibits an EP while the gauging covariance is intrinsically anisotropic
and phase sensitive, cleanly separating EP location (drift) from noise imprint ($S$). We then turn to explicitly
non-Markovian families of Gaussian channels, where EP manifolds are obtained from defectiveness of $X_t$ at fixed time and
the corresponding Stein solution $S_t$ can be computed analytically, including directly on the EP set. We conclude by
using the drift-controlled criterion to clarify why common phase-insensitive damping channels are EP-free, and to
highlight minimal non-normal drift mechanisms (e.g.\ critical damping) where EPs and Jordan-block transients arise
naturally.

Finally, we conclude in~\Cref{sec:conclusions} with a summary of our results and a discussion of future perspectives.

\section{Bosonic Gaussian channels}
\label{sec:gaussian_channels}

We consider $N$ bosonic modes with quadratures ordered as
\begin{equation}
\hat x := (\hat q_1,\dots,\hat q_N,\hat p_1,\dots,\hat p_N)^{\T}\in\mathbb{R}^{2N},
\label{eq:x_def}
\end{equation}
satisfying the canonical commutation relations
\begin{equation}
[\hat x,\hat x^{\T}] = \ii\,\Sigma,\quad
\Sigma := \begin{pmatrix} 0 & I \\ -I & 0 \end{pmatrix}.
\label{eq:CCR_Sigma}
\end{equation}
We define Weyl operators $W(\xi):=\exp\!\big(\ii\,\xi^{\T}\Sigma \hat x\big)$ and the Weyl (symmetrically ordered) characteristic function
\begin{equation}
\chi_\rho(\xi):=\Tr[\rho\,W(\xi)].
\label{eq:chi_def}
\end{equation}
The first-moment vector and centered quadratures are
\begin{equation}
d:=\langle \hat x\rangle_\rho=\Tr(\rho\,\hat x),
\quad
\Delta \hat x := \hat x - d,
\label{eq:d_and_Deltax}
\end{equation}
and the (symmetrized) covariance matrix is
\begin{align}
V
&:= \frac12\,\big\langle \{\Delta \hat x,(\Delta \hat x)^{\T}\}\big\rangle_\rho
= \frac12\,\Tr\!\Big[\rho\,\{\Delta \hat x,(\Delta \hat x)^{\T}\}\Big].
\label{eq:cov_def}
\end{align}

\subsection{Gaussian channels and $(X,Y,\delta)$ parametrization}
\label{subsec:Gaussian_XYdelta}

A Gaussian channel $\Psi$ is specified by the triple $(X,Y,\delta)$, with $X\in\mathbb{R}^{2N\times 2N}$,
$Y=Y^{\T}\in\mathbb{R}^{2N\times 2N}$, and $\delta\in\mathbb{R}^{2N}$, acting on the characteristic function as~\cite{CarusoGiovannettiHolevo2014}
\begin{equation}
\chi_{\Psi(\rho)}(\xi)
=
\exp\!\Big(-\tfrac12\,\xi^{\T}Y\,\xi + \ii\,\delta^{\T}\xi\Big)\;
\chi_\rho(X^{\T}\xi).
\label{eq:GaussianChannelChi}
\end{equation}
Equivalently, on first and second moments,
\begin{equation}
d \mapsto X d + \delta,\quad 
V \mapsto X V X^{\T} + Y.
\label{eq:GaussianChannelMoments}
\end{equation}
Complete positivity is equivalent to
\begin{equation}
Y + \frac{\ii}{2}\Big(\Sigma - X\Sigma X^{\T}\Big)\succeq 0,
\label{eq:CPconstraint}
\end{equation}
where $\succeq$ denotes the L\"owner (semidefinite) order: for a Hermitian matrix $M$, $M\succeq 0$ means that $M$ is
positive semidefinite (all eigenvalues are nonnegative), equivalently $v^\dagger M v\ge 0$ for all vectors $v$.

The parametrization \eqref{eq:GaussianChannelChi} is closed under composition. If
\[
\Psi_1\equiv(X_1,Y_1,\delta_1),\quad \Psi_2\equiv(X_2,Y_2,\delta_2),
\]
then $\Psi_{21}:=\Psi_2\circ\Psi_1$ is Gaussian with parameters~\cite{CarusoGiovannettiHolevo2014}
\begin{equation}
(X_{21},Y_{21},\delta_{21})
=
\bigl(X_2X_1,\;X_2Y_1X_2^{\T}+Y_2,\;X_2\delta_1+\delta_2\bigr).
\label{eq:Gaussian_composition_rule}
\end{equation}
This follows directly by inserting \eqref{eq:GaussianChannelChi} twice and collecting Gaussian prefactors; a detailed
proof (via characteristic functions and, equivalently, via moments) is given in \Cref{app:composition_rule}.

\subsection{Linear Lindbladians and Gaussian diffusion generators}
\label{subsec:linearL_to_AD}

Following Ma--Woolley--Petersen~\cite{MaWoolley2018}, we connect the standard Lindblad description of a linear open quantum system—i.e., a quadratic Hamiltonian with Lindblad operators linear in the quadratures $\hat x$—to the diffusion-generator data $(A,D,u)$ that govern the evolution of the characteristic function for an arbitrary quantum state.

Consider the Markovian master equation
\begin{equation}
\dot\rho
=
-\ii[\hat H,\rho]
+\sum_{j=1}^{K}\Big(\hat L_j\rho\,\hat L_j^\dagger-\tfrac12\{\hat L_j^\dagger \hat L_j,\rho\}\Big),
\label{eq:GKSL_linear}
\end{equation}
with quadratic Hamiltonian (allowing linear driving)
\begin{equation}
\hat H=\tfrac12\,\hat x^{\T}H\,\hat x - f^{\T}\hat x,
\quad H=H^{\T}\in\mathbb R^{2N\times 2N},\ \ f\in\mathbb R^{2N},
\label{eq:H_quad_linear}
\end{equation}
and Lindblad operators linear in $\hat x$,
\begin{equation}
\hat L_j=\ell_j^{\T}\hat x,
\quad \ell_j\in\mathbb C^{2N\times 1},\quad j\in\{1,\,\dots,\,K\}.
\label{eq:L_linear}
\end{equation}
The induced first- and second-moment dynamics closes and takes the diffusive form
\begin{equation}
\dot d(t)=A\,d(t)+u,\quad
\dot V(t)=A\,V(t)+V(t)\,A^{\T}+D,
\label{eq:momentODEs_abs}
\end{equation}
with drift and diffusion matrices \cite{MaWoolley2018}
\begin{align}
A&=\Sigma\Big(H+\Imm\!\big(C^\dagger C\big)\Big),
\quad
D=\Sigma\,\Ree\!\big(C^\dagger C\big)\,\Sigma^{\T},
\label{eq:AD_from_GKSL}
\\
C&:=(\ell_1,\dots,\ell_K)^{\T} \in \mathbb C^{K\times 2N},
\quad
C^\dagger C= \sum_{j=1}^{K}\ell_j^*\,\ell_j^\T,
\nonumber
\end{align}
and constant drift vector inherited from the linear Hamiltonian term,
\begin{equation}
u=\Sigma f.
\label{eq:u_from_drive}
\end{equation}
When $f=0$ one recovers the homogeneous mean equation $\dot d=A d$ emphasized in Ref.~\cite{MaWoolley2018}.

The pair $(A,D)$ defines a legitimate (completely positive) Gaussian Markov generator iff \cite{SerafiniBook}
\begin{equation}
D+\frac{\ii}{2}\big(A\Sigma+\Sigma A^{\T}\big)\succeq 0,
\label{eq:CP_generator_constraint_abs}
\end{equation}
the infinitesimal analogue of \eqref{eq:CPconstraint}.

\subsection{Ornstein--Uhlenbeck form and continuous-time Lyapunov gauging}
\label{subsec:OU_and_gauging}

The generator associated with \eqref{eq:momentODEs_abs} admits a phase-space representation on characteristic functions.
Writing $\chi(\xi,t):=\chi_{\rho(t)}(\xi)$, one obtains the Ornstein--Uhlenbeck evolution~\cite{SerafiniBook}
\begin{align}
    & \partial_t \chi(\xi,t) = \mathcal L\,\chi(\xi,t)\nonumber\\
    & := -\tfrac12\,\xi^{\T}D\,\xi\;\chi(\xi,t) +\big(A^{\T}\xi\big)\!\cdot\nabla_\xi \chi(\xi,t)
    +\ii\,u^{\T}\xi\;\chi(\xi,t),
\label{eq:OU_generator_general}
\end{align}
where ${\mathcal L}$ is the Ornstein--Uhlenbeck generator.
Assume the drift is stable (Hurwitz),
\begin{equation}
\Ree\,\lambda(A)<0.
\label{eq:Hurwitz_abs}
\end{equation}
Then the continuous Lyapunov equation
\begin{equation}
A S + S A^{\T}+D=0
\label{eq:Lyapunov_abs}
\end{equation}
admits a unique real symmetric solution $S=S^{\T}$, with~\cite{KhalilNonlinearSystems}
\begin{equation}
S=\int_{0}^{\infty}\e^{At}\,D\,\e^{A^{\T}t}\,\dd t.
\label{eq:Lyapunov_integral_solution}
\end{equation}
See~\cref{app:Lyapunov_integral_derivation} for a derivation of this result.

\begin{theorem}[Continuous-time (Lyapunov) diffusion gauging]
\label{thm:ct_gauging}
Assume \eqref{eq:Hurwitz_abs} and let $S$ solve \eqref{eq:Lyapunov_abs}. Define
\begin{equation}
\widetilde\chi(\xi,t):=\exp\!\big(\tfrac12\xi^{\T}S\xi\big)\chi(\xi,t).
\label{eq:chi_similarity_def}
\end{equation}
Then \eqref{eq:OU_generator_general} is similar to the diffusion-free drift generator
\begin{equation}
\partial_t \widetilde\chi(\xi,t)
=
\big(A^{\T}\xi\big)\!\cdot\nabla_\xi \widetilde\chi(\xi,t)
+\ii\,u^{\T}\xi\;\widetilde\chi(\xi,t).
\label{eq:OU_gauged_generator}
\end{equation}
Equivalently, at the channel level the semigroup $\{\Psi_t\}_{t\ge0}$ is similar, for every fixed $t\ge0$, to a Gaussian
channel with the same drift $X_t=\e^{At}$ and zero diffusion.
\end{theorem}

\begin{proof}
We first derive \eqref{eq:OU_gauged_generator}. By the product rule,
\begin{align}
\partial_t\widetilde\chi
&=
\exp\!\big(\tfrac12\xi^{\T}S\xi\big)\,\partial_t\chi,
\nonumber\\
\nabla_\xi \chi
&=
\exp\!\big(-\tfrac12\xi^{\T}S\xi\big)\Big(\nabla_\xi\widetilde\chi - S\xi\,\widetilde\chi\Big).
\label{eq:grad_transform}
\end{align}
Substituting \eqref{eq:OU_generator_general} and \eqref{eq:grad_transform} gives
\begin{align}
\partial_t \widetilde\chi
&=
-\tfrac12\,\xi^{\T}D\,\xi\;\widetilde\chi
+\big(A^{\T}\xi\big)\!\cdot\Big(\nabla_\xi \widetilde\chi - S\xi\,\widetilde\chi\Big)
+\ii\,u^{\T}\xi\;\widetilde\chi
\nonumber\\
&=
\big(A^{\T}\xi\big)\!\cdot\nabla_\xi \widetilde\chi
+\Big(-\tfrac12\,\xi^{\T}D\,\xi - \xi^{\T}A S\xi\Big)\widetilde\chi
+\ii\,u^{\T}\xi\;\widetilde\chi.
\label{eq:OU_intermediate}
\end{align}
\noindent\emph{Notation.} Here ``$\cdot$'' denotes the standard Euclidean pairing between a vector field and the gradient operator:
\begin{align} 
    &(A^{\T}\xi)\cdot\nabla_\xi:=\sum_{j=1}^{2N}(A^{\T}\xi)_j\,\partial_{\xi_j}.
\end{align}
In particular,
\begin{align} 
    &(A^{\T}\xi)\cdot(S\xi)=(A^{\T}\xi)^{\T}(S\xi)=\xi^{\T}AS\,\xi.
\end{align}
Using symmetry of $S$ and
\begin{equation}
\xi^{\T}(AS)\xi=\tfrac12\,\xi^{\T}\big(AS+SA^{\T}\big)\xi,
\label{eq:quadratic_symmetrization}
\end{equation}
the quadratic term becomes
$-\tfrac12\,\xi^{\T}\big(D+AS+SA^{\T}\big)\xi\,\widetilde\chi$ and vanishes by \eqref{eq:Lyapunov_abs}, yielding
\eqref{eq:OU_gauged_generator}.

To connect to the Gaussian channel parametrization, let $(X_t,Y_t,\delta_t)$ denote the Gaussian parameters of $\Psi_t$. Solving the moment equations \eqref{eq:momentODEs_abs} and comparing to \eqref{eq:GaussianChannelMoments} gives
\begin{align}
X_t&=\e^{At},\quad
\delta_t=\int_{0}^{t}\e^{A(t-s)}u\,\dd s,\quad
Y_t=\int_{0}^{t}\e^{As}D\,\e^{A^{\T}s}\,\dd s.
\label{eq:semigroup_params_recall}
\end{align}
    Define $\mathcal V_S:=(I,S,0)$, with formal inverse ${\mathcal V_S^{-1}\equiv (I,-S,0)}$.
    Gaussian composition implies, for each fixed $t$,
\begin{equation}
\mathcal V_S^{-1}\circ(X_t,Y_t,\delta_t)\circ\mathcal V_S
=
\bigl(X_t,\;Y_t+X_t S X_t^{\T}-S,\;\delta_t\bigr).
\label{eq:tildeY_general_abs}
\end{equation}
Thus diffusion is removed for all $t\ge0$ iff $Y_t=S-X_t S X_t^{\T}$ for all $t$. Differentiating this identity at $t=0$
and using $\dot X_0=A$ yields $\dot Y_0=-(AS+SA^{\T})$. Since $\dot Y_0=D$ from \eqref{eq:semigroup_params_recall}, we
recover \eqref{eq:Lyapunov_abs}. Conversely, if \eqref{eq:Lyapunov_abs} holds, then $F(t):=S-X_t S X_t^{\T}$ and $Y_t$
solve the same linear ODE $\dot Z=A Z+Z A^{\T}+D$ with $Z(0)=0$, hence coincide by uniqueness. Substituting into
\eqref{eq:tildeY_general_abs} gives $\widetilde Y_t\equiv0$ for all $t\ge0$.
\end{proof}

\paragraph*{Drift control of spectrum and Jordan structure.}
Assume $A$ is Hurwitz so that the Lyapunov equation admits a (unique) symmetric solution $S=S^{\T}$.
The Lyapunov transform \eqref{eq:chi_similarity_def} corresponds to an invertible change of representation,
\begin{equation}
\widetilde{\mathcal L}=T\,\mathcal L\,T^{-1},
\qquad
(Tf)(\xi):=\exp\!\big(\tfrac12\xi^{\T}S\xi\big)\,f(\xi),
\label{eq:similarity_maintext}
\end{equation}
under which the Ornstein--Uhlenbeck generator \eqref{eq:OU_generator_general} is mapped to the drift-only first-order
operator \eqref{eq:OU_gauged_generator}. Because $T$ is invertible on the natural class of test functions used in
\Cref{app:drift_controls_spectrum}, this transformation preserves eigenvalues and Jordan-block sizes. In particular, the
locations of exceptional points are fixed by the drift matrix $A$: the diffusion $D$ affects eigenfunctions through the
dressing covariance $S$, but it does not change the spectrum or the Jordan structure. A precise statement and proof are
given in \Cref{app:drift_controls_spectrum}.

When $u=0$ and $A$ is diagonalizable, this dependence on $A$ can be made completely explicit. Writing
$A^{\T}=V\Lambda V^{-1}$ with $\Lambda=\mathrm{diag}(\lambda_1,\ldots,\lambda_{2N})$ and introducing diagonalizing
coordinates $\eta:=V^{-1}\xi$, the gauged generator takes the Euler form
\begin{equation}
\widetilde{\mathcal L}=(A^{\T}\xi)\cdot\nabla_\xi
=\sum_{j=1}^{2N}\lambda_j\,\eta_j\,\partial_{\eta_j},
\label{eq:Ltilde_Euler_maintext}
\end{equation}
so monomials $m_{\mathbf n}(\eta):=\prod_{j}\eta_j^{n_j}$ are eigenfunctions with eigenvalue
$\sum_j n_j\lambda_j(A)$. Hence
\begin{equation}
\mathrm{spec}(\mathcal L)
=
\mathrm{spec}(\widetilde{\mathcal L})
=
\Big\{\;\sum_{j=1}^{2N} n_j\,\lambda_j(A)\ :\ n_j\in\mathbb N_0\;\Big\},
\label{eq:Liouvillian_spec_from_A_general}
\end{equation}
a standard result for Ornstein--Uhlenbeck generators that we re-derive in \Cref{app:additive_spectrum} (see also
Ref.~\cite{MetafunePallaraPriola2002}).

More generally, if $A$ is not diagonalizable, then the drift part already forces non-diagonalizability of the
Liouvillian. Concretely, if $A^{\T}$ contains a Jordan block $A^{\T}=\lambda I+N$ with nilpotent $N\neq 0$, the drift
operator $\widetilde{\mathcal L}=((\lambda I+N)\xi)\cdot\nabla_\xi$ acts non-diagonalizably on finite-dimensional spaces
of homogeneous polynomials, producing Jordan chains whose length increases with the polynomial degree. This gives a
constructive mechanism for Liouvillian exceptional points that depends only on $A$; see \Cref{app:defective_A_implies_EP}.

\subsection{Linear evolution in white-noise Gaussian environments}
\label{subsec:white_noise_model}

It is often convenient to specify the triplet $(A,D,u)$ without explicit reference to Lindblad operators, by starting from
a microscopic system--bath model and taking the white-noise (Markov) limit. We consider a quadratic system Hamiltonian
with linear driving,
\begin{equation}
\hat H_S=\frac12\,\hat x^{\T}H_S\hat x-u^{\T}\hat x,
\quad
H_S=H_S^{\T}\in\mathbb R^{2N\times 2N},\ \ u\in\mathbb R^{2N},
\label{eq:HS_quadratic_drive}
\end{equation}
together with a Gaussian environment of $M$ bosonic modes with quadratures $\hat x_{\rm in}\in\mathbb R^{2M}$, symplectic
form $\Sigma_{\rm in}$, and quadratic bath Hamiltonian
\begin{equation}
\hat H_{\rm in}=\frac12\,\hat x_{\rm in}^{\T}H_{\rm in}\hat x_{\rm in},
\quad
H_{\rm in}=H_{\rm in}^{\T}\in\mathbb R^{2M\times 2M}.
\label{eq:Hbath_quad}
\end{equation}
We assume the bath is prepared in a stationary Gaussian state with covariance matrix
$\sigma_{\rm in}=\sigma_{\rm in}^{\T}\in\mathbb R^{2M\times 2M}$ and take a bilinear system--bath interaction
\begin{equation}
\hat H_{\rm int}=\hat x^{\T}C\,\hat x_{\rm in},
\quad
C\in\mathbb R^{2N\times 2M}.
\label{eq:Hint_bilinear}
\end{equation}
In the white-noise limit the reduced dynamics is Gaussian and takes the diffusive form~\eqref{eq:momentODEs_abs}; see
Ref.~\cite{SerafiniBook} for details.

In this Markovian setting the drift and diffusion matrices read~\cite{SerafiniBook}
\begin{equation}
A=\Sigma H_S+\frac12\,\Sigma C\,\Sigma_{\rm in}\,C^{\T},
\quad
D=\Sigma\,C\,\sigma_{\rm in}\,C^{\T}\Sigma^{\T}.
\label{eq:AD_white_noise}
\end{equation}
The bath covariance must satisfy the Robertson--Schr\"odinger uncertainty principle,
\begin{equation}
\sigma_{\rm in}+\frac{\ii}{2}\Sigma_{\rm in}\succeq 0,
\label{eq:env_RS_constraint}
\end{equation}
which implies the generator-level complete-positivity constraint~\eqref{eq:CP_generator_constraint_abs}. In particular,
the Lyapunov solution $S$ in~\eqref{eq:Lyapunov_abs} depends explicitly on $H_S$, $C$, and the bath state through
$\sigma_{\rm in}$ (and on $\Sigma_{\rm in}$ through $A$).

Reference~\cite{SerafiniBook} adopts the interleaved ordering
${\hat r\equiv (\hat q_1,\hat p_1,\ldots,\hat q_N,\hat p_N)^{\T}}$, with symplectic form
${\Omega\equiv \bigoplus_{j=1}^N \begin{pmatrix} 0 & 1 \\ -1 & 0 \end{pmatrix}}$, whereas we use the grouped ordering
    ${\hat x\equiv(\hat q_1,\ldots,\hat q_N,\hat p_1,\ldots,\hat p_N)^{\T}}$, with symplectic form
${\Sigma\equiv \begin{pmatrix} 0 & \mathbbm 1 \\ -\mathbbm 1 & 0 \end{pmatrix}}$.
    The Markov-limit expressions retain the
same algebraic form under the corresponding permutation. Concretely, let $P\in\mathbb R^{2N\times 2N}$ be the permutation
matrix such that $x=Pr$, i.e.
\begin{equation}
P_{i,\,2i-1}=1,\quad P_{N+i,\,2i}=1\quad (i=1,\ldots,N),
\label{eq:perm_matrix_P}
\end{equation}
with all other entries zero. Then $P^{-1}=P^{\T}$ and
\begin{align}
    & \Sigma = P\,\Omega\,P^{\T},
\quad
\sigma_x = P\,\sigma_r\,P^{\T},\nonumber\\
    & A_x = P\,A_r\,P^{\T},
\quad
D_x = P\,D_r\,P^{\T},
\label{eq:perm_conjugation_rules}
\end{align}
and analogously for the input matrices if their quadratures are reordered.
Thus the formulas in Ref.~\cite{SerafiniBook} coincide with Eqs.~\eqref{eq:AD_white_noise}--\eqref{eq:env_RS_constraint} once expressed in a common convention.

\subsection{From generators to channels}
\label{subsec:generator_to_channel_and_beyond}

Theorem~\ref{thm:ct_gauging} has an immediate channel-level interpretation: for a Markov Gaussian semigroup
$\Psi_t\equiv(X_t,Y_t,\delta_t)$ with $X_t=\e^{At}$ and
$Y_t=\int_0^t \e^{As}D\,\e^{A^{\T}s}\dd s$, the similarity transform by the Gaussian smoothing map
$\mathcal V_S=(I,S,0)$ removes diffusion \emph{for all} $t\ge 0$,
\begin{equation}
\mathcal V_S^{-1}\circ (X_t,Y_t,\delta_t)\circ \mathcal V_S
=
\bigl(X_t,\,0,\,\delta_t\bigr),
\quad \forall\,t\ge 0,
\end{equation}
if and only if $S$ solves the Lyapunov equation $AS+SA^{\T}+D=0$. In particular, within a semigroup the diffusion-gauging
condition is equivalent whether stated at the generator level (Lyapunov) or at the channel level (vanishing $Y_t$ after
similarity).

This continuous-time construction motivates an analogous \emph{discrete-time} statement for a \emph{single} stable Gaussian
channel $\Psi\equiv(X,Y,\delta)$. Importantly, this is not a corollary of Theorem~\ref{thm:ct_gauging}: a generic channel
need not be embeddable into a Markov semigroup $\Psi_t=\exp(t\mathcal L)$, and the discrete-time gauging condition is a
single-map statement governed by a Stein equation rather than a Lyapunov equation.

\subsection{Noise gauging for a Gaussian channel}
\label{subsec:discrete_gauging}

We begin by recalling that displacements can be removed by unitary conjugation. The deterministic displacement channel
$\Phi_\chi(\rho)=W(\chi)\rho W(\chi)^\dagger$ has Gaussian parameters $(I,0,\chi)$ and satisfies
$\Phi_\chi^\dagger=\Phi_{-\chi}$. Conjugating $\Psi\equiv(X,Y,\delta)$ by $\Phi_\chi$ yields \cite{Weedbrook2012}
\begin{align}
\Psi_\chi &:= \Phi_\chi \circ \Psi \circ \Phi_\chi^\dagger,\\
(X_\chi,Y_\chi,\delta_\chi) &= \bigl(X,\;Y,\;\delta + (I-X)\chi\bigr).
\label{eq:displacementConjugation_app}
\end{align}
Thus, whenever $(I-X)$ is invertible, choosing ${\chi=-(I-X)^{-1}\delta}$ sets $\delta_\chi=0$ without affecting the
spectral data encoded in $X$.

To gauge diffusion we introduce the Gaussian smoothing map
\begin{equation}
\mathcal V_S := (I,\;S,\;0),
\quad S=S^{\T}\in\mathbb R^{2N\times 2N},
\label{eq:VS_channel_params}
\end{equation}
together with its formal inverse $\mathcal V_S^{-1}\equiv(I,-S,0)$. We use $\mathcal V_S$ as a similarity
transformation at the level of Gaussian parameters.

Assume the channel is \emph{stable},
\begin{equation}
\spr(X)<1.
\label{eq:stability_discrete_app}
\end{equation}
Then the (discrete-time) Stein equation \cite{Simoncini2016}
\begin{equation}
S = X S X^{\T} + Y
\label{eq:Stein_app}
\end{equation}
admits a unique real symmetric solution. Moreover, it has the convergent series representation
\begin{equation}
S=\sum_{n=0}^{\infty} X^{n}Y\big(X^{\T}\big)^{n},
\label{eq:SteinSeries_general_app}
\end{equation}
which immediately implies that $S\succeq 0$ whenever $Y\succeq 0$.

\begin{theorem}[Discrete-time (Stein) noise gauging]
\label{thm:dt_gauging}
Let $\Psi\equiv(X,Y,\delta)$ satisfy \eqref{eq:stability_discrete_app} and let $S$ solve \eqref{eq:Stein_app}. Then the
similarity-transformed map
\begin{equation}
\widetilde\Psi := \mathcal V_S^{-1}\circ \Psi \circ \mathcal V_S
\label{eq:similarity_def_app}
\end{equation}
has Gaussian parameters
\begin{equation}
\widetilde\Psi \equiv (X,\;0,\;\delta).
\label{eq:gaugeAwayYchannel_app}
\end{equation}
In particular, $\Psi$ and $\widetilde\Psi$ are isospectral and share the same Jordan structure.
\end{theorem}

\begin{proof}
Gaussian composition gives
\begin{equation}
\mathcal V_S^{-1}\circ(X,Y,\delta)\circ\mathcal V_S
=
\bigl(X,\;Y+X S X^{\T}-S,\;\delta\bigr).
\label{eq:tilde_params_preStein}
\end{equation}
Imposing \eqref{eq:Stein_app} is equivalent to $Y=S-XSX^{\T}$, which sets $\widetilde Y=0$ in
\eqref{eq:tilde_params_preStein}; the identities $\widetilde X=X$ and $\widetilde\delta=\delta$ follow directly.
\end{proof}

\begin{remark*}[Gauging away $Y$ does \emph{not} force $X$ to be symplectic]
\label{rem:gauging_not_CP}
It is tempting to combine the gauged form $\widetilde\Psi\equiv(X,0,\delta)$ with the complete-positivity (CP) condition
for Gaussian channels,
\begin{equation}
Y+\frac{i}{2}\bigl(\Sigma-X\Sigma X^{\T}\bigr)\succeq 0,
\label{eq:CP_constraint_rem}
\end{equation}
and infer that setting $Y=0$ forces $X$ to be symplectic. Indeed, if one demands that the \emph{diffusion-free} map
$(X,0,\delta)$ itself be a physical CPTP Gaussian channel, then \eqref{eq:CP_constraint_rem} becomes
$\tfrac{i}{2}(\Sigma-X\Sigma X^{\T})\succeq 0$. Since $\Sigma-X\Sigma X^{\T}$ is real antisymmetric,
$i(\Sigma-X\Sigma X^{\T})$ is Hermitian with spectrum symmetric about $0$, so it can be positive semidefinite only if it
vanishes. Hence CP would enforce $\Sigma=X\Sigma X^{\T}$ and $X\in\Sp(2N,\mathbb R)$.

There is no contradiction with Theorem~\ref{thm:dt_gauging}, because the noise gauging is a \emph{similarity
transformation}, not a CPTP conjugation by physical channels. While $\mathcal V_S\equiv(I,S,0)$ is CP whenever $S\succeq 0$,
its formal inverse $\mathcal V_S^{-1}\equiv(I,-S,0)$ is generically \emph{not} CP (unless $S\preceq 0$). Consequently the
transformed map
\[
\widetilde\Psi=\mathcal V_S^{-1}\circ\Psi\circ\mathcal V_S
\]
need not satisfy the CP inequality \eqref{eq:CP_constraint_rem}. The role of $\mathcal V_S^{\pm 1}$ is instead to
implement an invertible change of representation (e.g.\ on characteristic functions), under which the associated
superoperator is similar. Therefore $\Psi$ and $\widetilde\Psi$ are isospectral and share the same Jordan structure, even
though $\widetilde\Psi$ is, in general, not a bona fide quantum channel. In particular, the gauging leaves $X$ invariant
and does \emph{not} enforce $X$ to be symplectic.
\end{remark*}

\subsection{Summary: implications for exceptional points}
\label{subsec:EP_summary}

For Gaussian Markov dynamics generated by \eqref{eq:momentODEs_abs}, the Ornstein--Uhlenbeck generator
\eqref{eq:OU_generator_general} is similar, under the Lyapunov transform \eqref{eq:chi_similarity_def}, to the diffusion-free
drift operator \eqref{eq:OU_gauged_generator}. Consequently, eigenvalues and Jordan structure and hence exceptional-point
manifolds are drift-controlled: they are determined by $A$, while diffusion $D$ modifies only eigenoperators and the
steady state through the dressing covariance $S$.

At the channel level, this statement is equivalent to uniform-in-time gauging of the semigroup diffusion matrix $Y_t$:
\eqref{eq:gaugeD_equiv_gaugeY}. Moreover, since $X_t=\exp(At)$ inherits the Jordan structure of $A$ for any fixed $t>0$,
searching for EPs of $A$ is equivalent to searching for EPs of $X_t$.
Motivated by this continuous-time structure, Theorem~\ref{thm:dt_gauging} shows that for a
single stable discrete-time Gaussian channel the same similarity mechanism gauges away $Y$ via the Stein solution, leaving
$X$ as the object controlling non-diagonalizability and exceptional points.

\section{Applications to single-mode Gaussian channels}
\label{sec:example_single_mode_nm}

We now specialize to one mode, where EPs correspond to a defective $2\times2$ drift. The gauging covariance provides a
compact way to track how noise enters without shifting EP locations: for a fixed-time map $\Psi_t\equiv(X_t,Y_t,\delta_t)$
the Stein solution $S_t$ is the covariance of the smoothing map $\mathcal V_{S_t}$ that removes $Y_t$ from the spectral
problem. Because $S_t$ is a real symmetric $2\times2$ matrix, its eigenvectors pick out the principal quadrature axes of
the required smoothing and its eigenvalues quantify the smoothing strength along those axes (whereas $\Tr S_t$ captures
only the aggregate amount).

We first treat a Markovian example (a degenerate parametric amplifier coupled to a squeezed reservoir) where the drift
has an EP while $S$ is phase sensitive, and then a non-semigroup family where EP lines follow from defectiveness of $X_t$
and $S_t$ remains analytically tractable on the EP set.

\subsection{A Markovian example: Single-mode squeezed-reservoir Lindbladian}
\label{subsec:squeezed_reservoir_EP}

We focus on a standard single-mode quantum--optical setting: a cavity mode undergoing linear damping into a broadband
\emph{squeezed} reservoir. This model is routinely realized by injecting squeezed vacuum into an optical cavity to achieve
phase-sensitive noise suppression, and likewise in microwave/circuit-QED platforms where a Josephson parametric amplifier
provides a squeezed input field that acts as an effectively Markovian bath over the relevant bandwidth
\cite{GardinerZoller,Wallsmilburn,Weedbrook2012,SerafiniBook}. In the Markov limit, the reduced dynamics is given by a
quadratic (parametric) Hamiltonian and a Bogoliubov jump operator \cite{GardinerZoller,SerafiniBook}.

From our perspective, this setting is useful because it simultaneously displays two features central to the present work.
First, the drift matrix associated with the linearized equations of motion can become \emph{defective}, producing an
exceptional point (EP) controlled entirely by the coherent/parametric part of the dynamics. Second, because the reservoir
is squeezed, the diffusion matrix is phase sensitive, and the corresponding Lyapunov-gauge covariance $S$ is generically
\emph{nontrivial} (anisotropic and, for generic squeezing phase, nondiagonal) \cite{Weedbrook2012,SerafiniBook}. This
provides a clean example where EP locations are fixed by the drift, while diffusion leaves a distinct, physically
meaningful imprint on the gauging covariance required to remove noise from the eigenvalue problem.

We take $\hat a=(\hat q+\ii \hat p)/\sqrt{2}$ and $\hat x=(\hat q,\hat p)^{\T}$ with
${\Sigma=\begin{pmatrix}0&1\\-1&0\end{pmatrix}}$, and consider the master equation
\begin{equation}
\dot\rho
=
-\ii[\hat H_S,\rho]
+
\kappa\,\mathcal D[\hat L]\rho,
\quad
\mathcal D[\hat L]\rho := \hat L\rho \hat L^\dagger -\tfrac12\{\hat L^\dagger \hat L,\rho\}.
\label{eq:ME_squeezed_res}
\end{equation}
The Hamiltonian is the degenerate parametric amplifier (up to the irrelevant constant ${-\frac{\Delta}{2}}$),
\begin{align}
\hat H_S
&=
\Delta\,\hat a^\dagger \hat a + \frac{\epsilon}{2}\big(\hat a^2+\hat a^{\dagger 2}\big)
=
\frac12\,\hat x^{\T}H_S \hat x,
\\
H_S
&=
\begin{pmatrix}
\Delta+\epsilon & 0\\
0 & \Delta-\epsilon
\end{pmatrix},
\label{eq:HS_DPA}
\end{align}
and the jump operator is a squeezed-reservoir (Bogoliubov) mode,
\begin{equation}
\hat L
=
\cosh r\,\hat a + e^{\ii\phi}\sinh r\,\hat a^\dagger,
\label{eq:L_squeezed_jump}
\end{equation}
with $r\ge 0$ the reservoir squeezing and $\phi$ its angle.

Writing $\hat L=c^{\T}\hat x$ with $c\in\mathbb C^2$, the coefficients follow directly from
$\hat a=(\hat q+\ii\hat p)/\sqrt{2}$ and $\hat a^\dagger=(\hat q-\ii\hat p)/\sqrt{2}$:
\begin{align}
\hat L
&=
\frac{1}{\sqrt{2}}\Big[(\cosh r+e^{\ii\phi}\sinh r)\hat q
+\ii(\cosh r-e^{\ii\phi}\sinh r)\hat p\Big],
\nonumber\\
c^{\T}
&=
\frac{1}{\sqrt{2}}\Big(\cosh r+e^{\ii\phi}\sinh r,\ \ \ii(\cosh r-e^{\ii\phi}\sinh r)\Big).
\label{eq:c_row}
\end{align}
In the Ma--Woolley--Petersen parametrization, for linear Lindblad operators $\hat L_j=(c_j)^{\T}\hat x$ one forms the
complex matrix $C$ whose rows are $c_j^{\T}$ and expresses drift and diffusion through $C^\dagger C$:
\begin{equation}
A = \Sigma\big(H_S+\Imm[C^\dagger C]\big),
\quad
D = \Sigma\,\Ree[C^\dagger C]\,\Sigma^{\T}.
\label{eq:AD_from_C_general}
\end{equation}
Here there is a single jump operator, hence ${C=c^{\T}\in\mathbb C^{1\times 2}}$.
A direct calculation (see~\Cref{app:explicit_cc}) yields
{\small
\begin{align}
& \Ree[C^\dagger C] \nonumber\\
&=
\frac{1}{2}
\begin{pmatrix}
\cosh(2r)+\sinh(2r)\cos\phi & \sinh(2r)\sin\phi\\
\sinh(2r)\sin\phi & \cosh(2r)-\sinh(2r)\cos\phi
\end{pmatrix},
\label{eq:Re_CdC}\\
& \Imm[C^\dagger C]
= \frac{1}{2}\,\Sigma.
\label{eq:Im_CdC}
\end{align}
}
Including the overall rate $\kappa$ in \eqref{eq:ME_squeezed_res} amounts to replacing
$C^\dagger C\mapsto \kappa\,C^\dagger C$. Substituting \eqref{eq:HS_DPA} and \eqref{eq:Re_CdC}--\eqref{eq:Im_CdC} into
\eqref{eq:AD_from_C_general} gives the drift
\begin{equation}
A
=
\Sigma H_S + \Sigma\Big(\frac{\kappa}{2}\Sigma\Big)
=
\begin{pmatrix}
-\kappa/2 & \Delta-\epsilon\\
-(\Delta+\epsilon) & -\kappa/2
\end{pmatrix},
\label{eq:A_DPA_squeezed}
\end{equation}
and the phase-sensitive diffusion matrix
{\small
\begin{align}
D
=
\frac{\kappa}{2}
\begin{pmatrix}
\cosh(2r)-\sinh(2r)\cos\phi & -\sinh(2r)\sin\phi\\
-\sinh(2r)\sin\phi & \cosh(2r)+\sinh(2r)\cos\phi
\end{pmatrix}.
\label{eq:D_squeezed}
\end{align}
}
Thus $D$ is generically non-isotropic and non-diagonal, so the Lyapunov-gauge covariance $S$ is also nontrivial.

The drift eigenvalues are
\begin{equation}
\lambda_{\pm}(A)= -\frac{\kappa}{2}\pm \sqrt{\epsilon^{2}-\Delta^{2}},
\label{eq:A_eigs}
\end{equation}
so $A$ becomes defective when the discriminant vanishes while $A$ is not proportional to the identity, i.e.\ on the EP
manifold
\begin{equation}
\Delta^2=\epsilon^2,
\quad \epsilon\neq 0,
\label{eq:EP_condition_DPA}
\end{equation}
which separates the oscillatory regime ($\Delta^2>\epsilon^2$) from the overdamped/unstable regime ($\Delta^2<\epsilon^2$).

Assume $A$ is Hurwitz, i.e.\ $\Ree\,\lambda_\pm(A)<0$. For the drift \eqref{eq:A_DPA_squeezed} this holds throughout the
stable region where $\epsilon^2\le \Delta^2$ (oscillatory regime) and also for part of the overdamped regime
$\epsilon^2>\Delta^2$ provided $\kappa>2\sqrt{\epsilon^2-\Delta^2}$.

To find the diffusion gauge $S$ we solve the Lyapunov equation
\begin{equation}
A S + S A^{\T} + D = 0,
\label{eq:Lyapunov_squeezed}
\end{equation}
with $S=S^{\T}$ and $D=D^{\T}$.

In the single-mode case ($2\times2$ matrices) the Lyapunov equation can be solved analytically in closed form. Write
\begin{align} \label{}
    &
A=\begin{pmatrix}a&b\\ c&d\end{pmatrix},\quad
S=\begin{pmatrix}s_{11}&s_{12}\\ s_{12}&s_{22}\end{pmatrix},\quad
D=\begin{pmatrix}d_{11}&d_{12}\\ d_{12}&d_{22}\end{pmatrix}.
\end{align}
Then $AS+SA^{\T}+D=0$ is equivalent to the $3\times3$ linear system (see~\cref{app:Lyapunov_vec_one_mode})
\begin{equation}
\begin{pmatrix}
2a & 2b & 0\\
c & a+d & b\\
0 & 2c & 2d
\end{pmatrix}
\begin{pmatrix}
s_{11}\\ s_{12}\\ s_{22}
\end{pmatrix}
=
-
\begin{pmatrix}
d_{11}\\ d_{12}\\ d_{22}
\end{pmatrix}.
\label{eq:Lyapunov_3x3_general}
\end{equation}

Solving the linear system \eqref{eq:Lyapunov_3x3_general_app} by explicit elimination gives the closed form
\begin{equation}
s_{12}
=
\frac{ab\,d_{22}+cd\,d_{11}-2ad\,d_{12}}
{2(a+d)(ad-bc)},
\label{eq:s12_Lyap_closed_app}
\end{equation}
whenever the denominator is nonzero (in particular, for Hurwitz $A$ one typically has $a+d\neq 0$ and $ad-bc\neq 0$).
The remaining entries follow directly from the first and third equations of
\eqref{eq:Lyapunov_3x3_general_app},
\begin{equation}
s_{11}
=
-\frac{d_{11}+2b\,s_{12}}{2a},
\quad
s_{22}
=
-\frac{d_{22}+2c\,s_{12}}{2d},
\label{eq:s11s22_from_s12_app}
\end{equation}
with the understanding that these expressions extend by continuity in nongeneric cases (e.g.\ $a=0$ or $d=0$) by solving
the $3\times3$ system directly.

For the drift matrix in Eq.~\eqref{eq:A_DPA_squeezed} one has
\[
a=d=-\kappa/2,\quad b=\Delta-\epsilon,\quad c=-(\Delta+\epsilon),
\]
and identifying $(s_{11},s_{12},s_{22})=(s_{qq},s_{qp},s_{pp})$ and $(d_{11},d_{12},d_{22})=(D_{qq},D_{qp},D_{pp})$,
Eqs.~\eqref{eq:s12_Lyap_closed_app}--\eqref{eq:s11s22_from_s12_app} reduce to
\begin{align}
s_{qp}
&=
\frac{\kappa D_{qp}+(\Delta-\epsilon)D_{pp}-(\Delta+\epsilon)D_{qq}}
{\kappa^{2}+4(\Delta^{2}-\epsilon^{2})},
\label{eq:sqp_general}\\[0.5ex]
s_{qq}
&=\frac{D_{qq}}{\kappa}+\frac{2(\Delta-\epsilon)}{\kappa}\,s_{qp},
\quad
s_{pp}
=\frac{D_{pp}}{\kappa}-\frac{2(\Delta+\epsilon)}{\kappa}\,s_{qp}.
\label{eq:sqq_spp_general}
\end{align}
Substituting the diffusion entries from \eqref{eq:D_squeezed} yields analytic expressions for
$S(\kappa,\Delta,\epsilon,r,\phi)$.

On the EP branch $\Delta=+\epsilon$ the denominator becomes $\kappa^{2}$ and the solution simplifies to
{\small
\begin{subequations}
\label{eq:S_EP_plus}
\begin{align}
s_{qq}\big|_{\Delta=\epsilon}
&=
\frac{1}{2}\Big(\cosh(2r)-\sinh(2r)\cos\phi\Big),
\\
s_{qp}\big|_{\Delta=\epsilon}
&=
-\frac{1}{2}\sinh(2r)\sin\phi
-\frac{\epsilon}{\kappa}\Big(\cosh(2r)-\sinh(2r)\cos\phi\Big),
\\
s_{pp}\big|_{\Delta=\epsilon}
&=
\frac{1}{2}\Big(\cosh(2r)+\sinh(2r)\cos\phi\Big)
+\frac{2\epsilon}{\kappa}\sinh(2r)\sin\phi\nonumber\\
&\quad +\frac{4\epsilon^{2}}{\kappa^{2}}\Big(\cosh(2r)-\sinh(2r)\cos\phi\Big),
\end{align}
\end{subequations}
}
and the other EP branch $\Delta=-\epsilon$ follows analogously (equivalently by $\epsilon\mapsto-\epsilon$).

The eigenvalues of $S$ quantify the Gaussian smoothing required by the similarity transform
$\mathcal V_S\equiv(I,S,0)$ to gauge away diffusion while preserving spectrum and Jordan structure. In this
squeezed-reservoir example, $S$ is generically anisotropic and phase-sensitive through $(r,\phi)$, while EP locations are
controlled entirely by defectiveness of the drift matrix $A$.

\subsubsection{Plots: EP locations and gauge eigenvalues vs.\ $\kappa$, $r$, $\phi$}

\begin{figure}[!htbp]
\centering
\includegraphics[width=0.9\columnwidth]{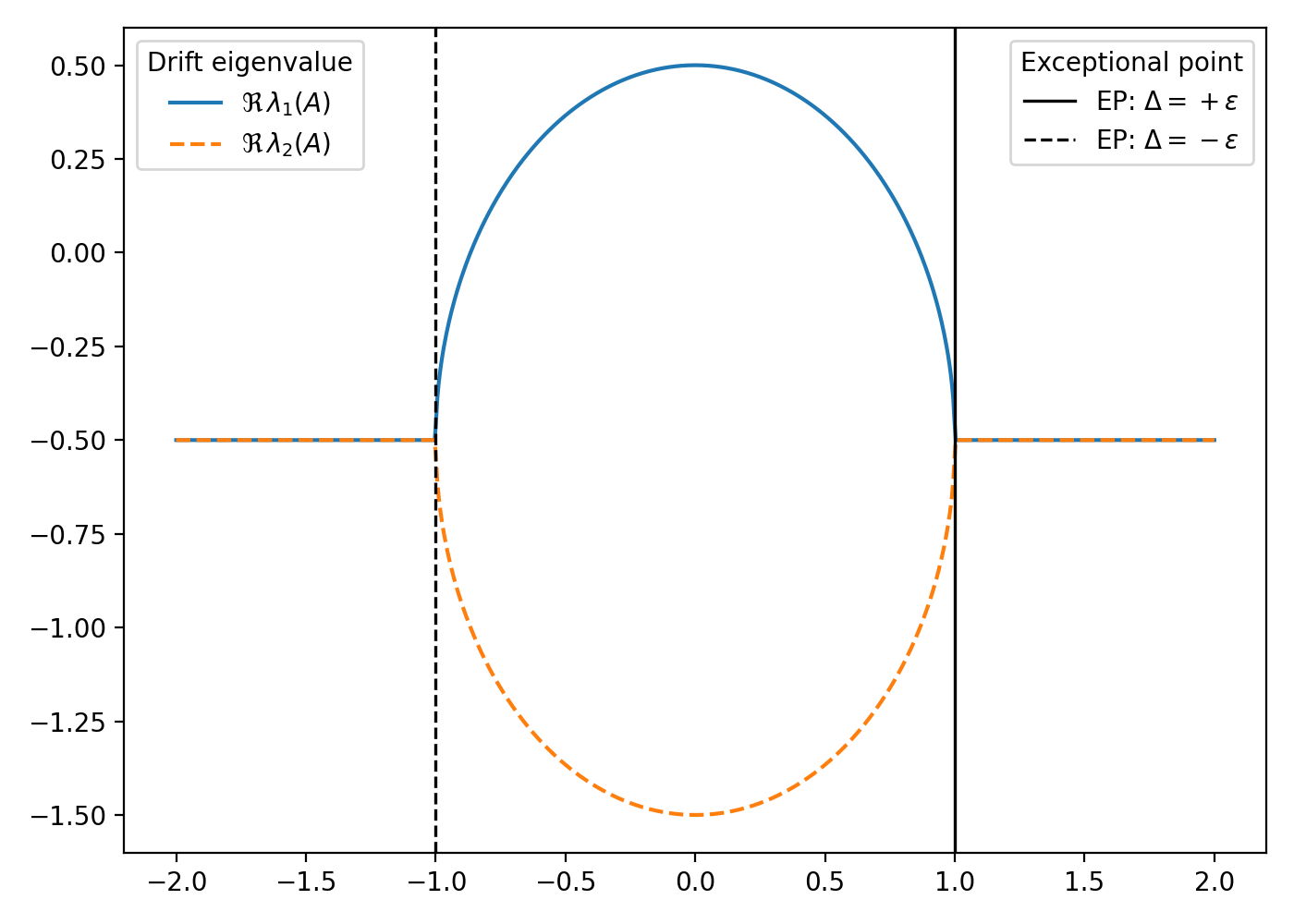}
\caption{\justifying Real parts of the drift eigenvalues $\lambda_{\pm}(A)$ vs.\ detuning $\Delta$ at fixed $(\kappa,\epsilon)$.
The EPs occur at $\Delta=\pm\epsilon$ where the square-root discriminant vanishes.}
\label{fig:drift_EP_locations}
\end{figure}

Figure~\ref{fig:drift_EP_locations} confirms that the drift spectrum exhibits the expected square-root coalescence:
the two eigenvalues $\lambda_\pm(A)$ merge at $\Delta=\pm\epsilon$, where the discriminant $\epsilon^2-\Delta^2$
vanishes and the drift becomes defective.

\begin{figure}[!htbp]
\centering
\begin{subfigure}[t]{0.9\columnwidth}
\centering
\includegraphics[width=\linewidth]{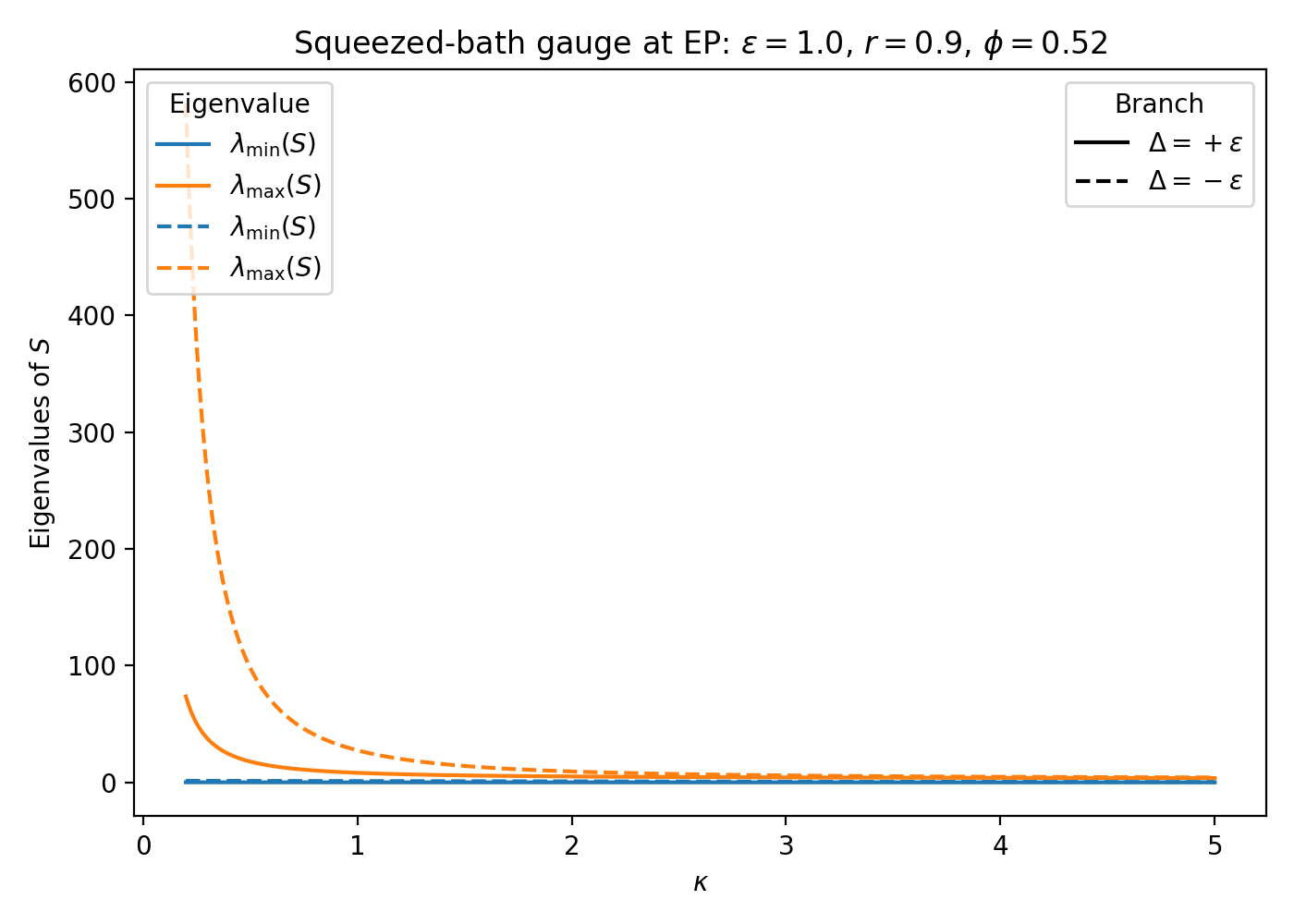}
\caption{\justifying Gauge eigenvalues $\lambda_{1,2}(S)$ along the EP branches $\Delta=\pm\epsilon$ versus damping rate $\kappa$, for fixed $(\epsilon,r,\phi)$.}
\label{fig:S_vs_kappa_EP}
\end{subfigure}

\vspace{0.6em}

\begin{subfigure}[t]{0.9\columnwidth}
\centering
\includegraphics[width=\linewidth]{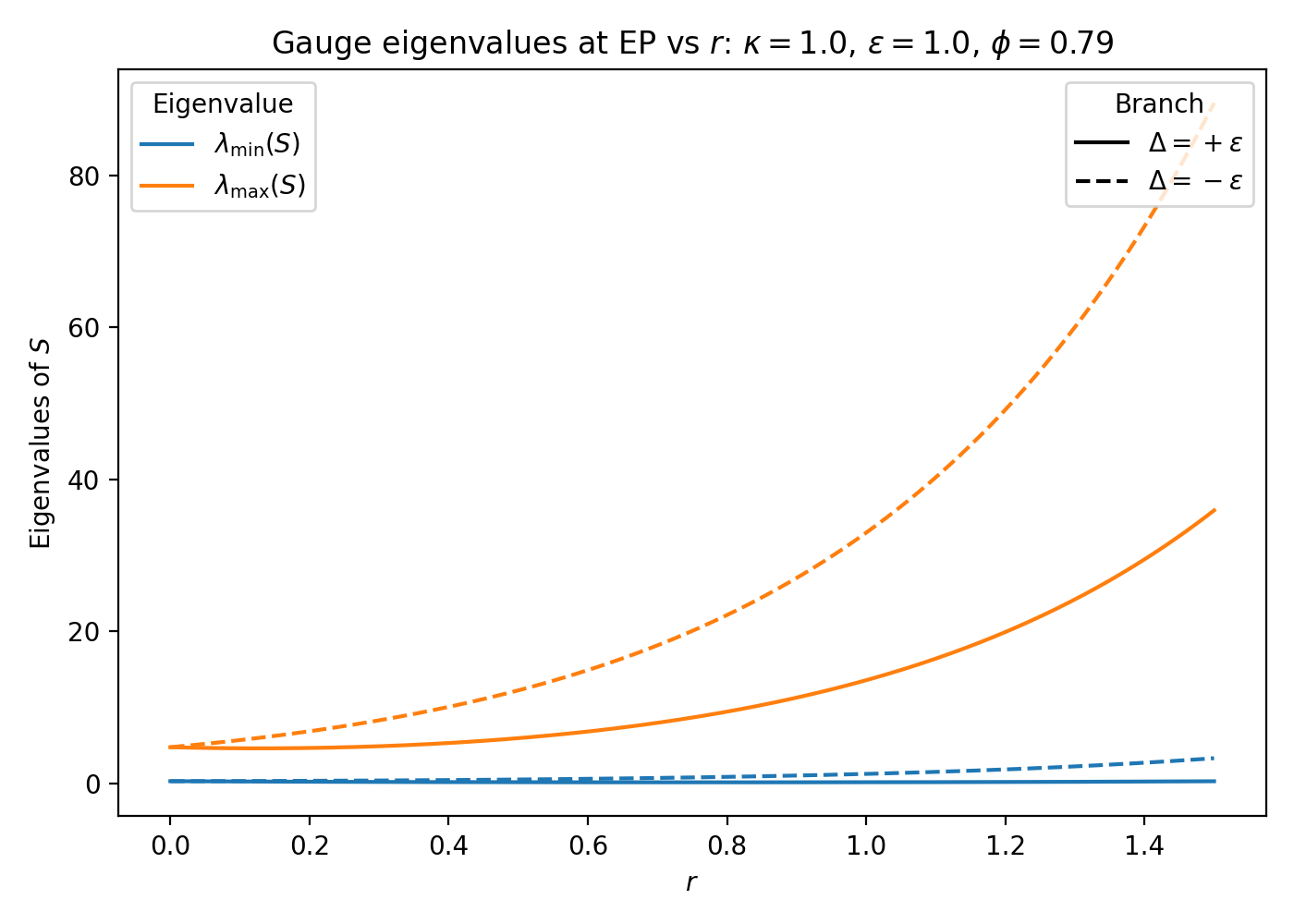}
\caption{\justifying Gauge eigenvalues $\lambda_{1,2}(S)$ along $\Delta=\pm\epsilon$ versus reservoir squeezing $r$, for fixed $(\kappa,\epsilon,\phi)$.}
\label{fig:S_vs_r_EP}
\end{subfigure}

\vspace{0.6em}

\begin{subfigure}[t]{0.9\columnwidth}
\centering
\includegraphics[width=\linewidth]{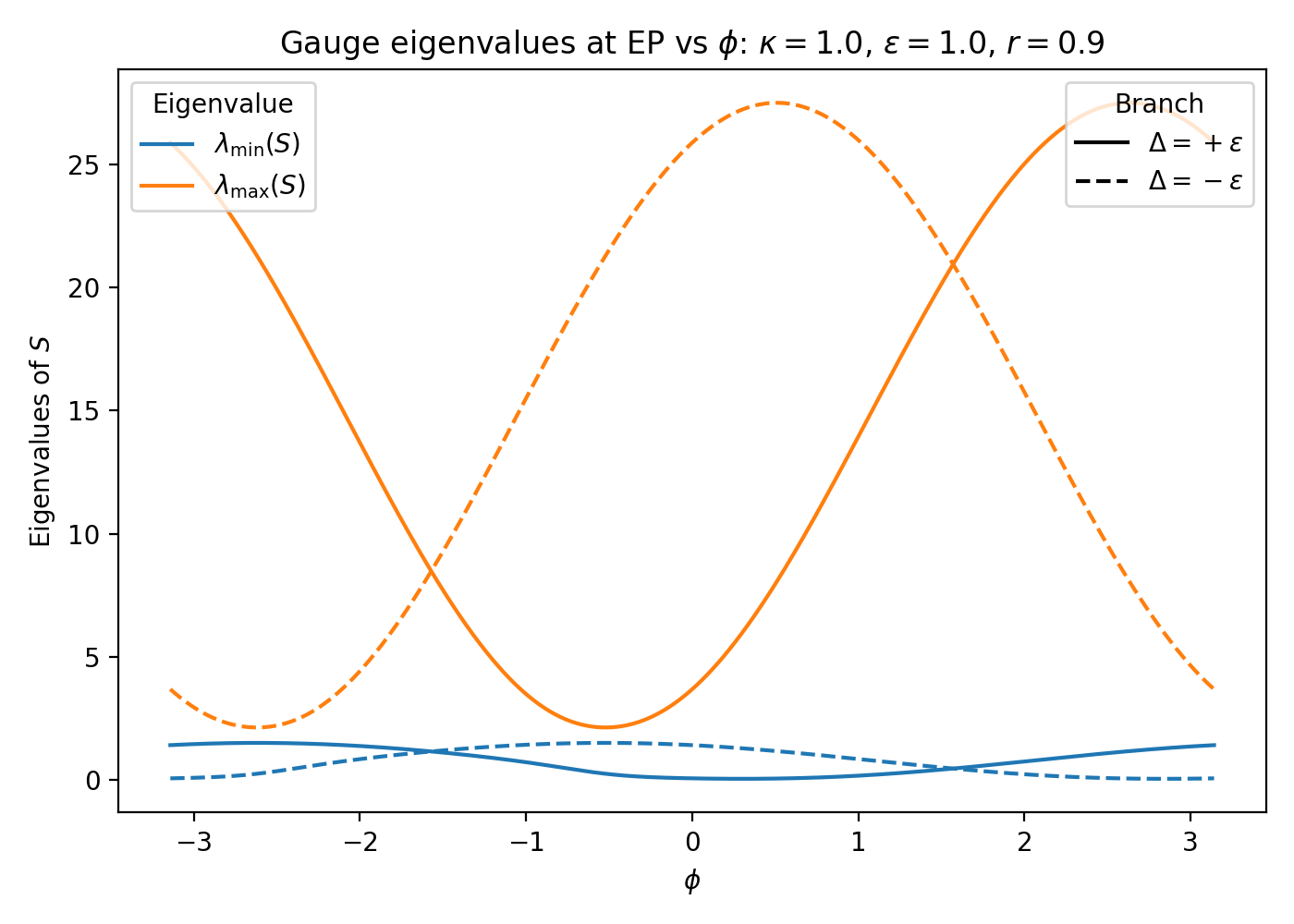}
\caption{\justifying Gauge eigenvalues $\lambda_{1,2}(S)$ along $\Delta=\pm\epsilon$ versus squeezing phase $\phi$, for fixed $(\kappa,\epsilon,r)$.}
\label{fig:S_vs_phi_EP}
\end{subfigure}

\caption{\justifying Eigenvalues of the Lyapunov gauge covariance $S$ on the EP branches $\Delta=\pm\epsilon$ for a squeezed-reservoir Lindbladian, shown as functions of (a) $\kappa$, (b) $r$, and (c) $\phi$ (with the remaining parameters held fixed as indicated).}
\label{fig:S_EP_triptych}
\end{figure}

Turning to the Lyapunov gauge, Fig.~\ref{fig:S_EP_triptych} resolves how the required gauging strength varies along
the EP branches. In Fig.~\ref{fig:S_vs_kappa_EP} the eigenvalues $\lambda_{1,2}(S)$ decrease with increasing damping
rate $\kappa$, indicating that stronger dissipation reduces the amount of Gaussian smoothing needed to gauge away the
diffusion (in the sense of the similarity transform generated by $S$). By contrast, Fig.~\ref{fig:S_vs_r_EP} shows that
$\lambda_{1,2}(S)$ grow rapidly with reservoir squeezing $r$: larger bath squeezing demands a larger gauging covariance,
consistent with the intuition that a more strongly phase-sensitive diffusion requires a stronger compensating gauge.

These monotonic trends with $\kappa$ and $r$ are qualitatively similar on both EP branches $\Delta=\pm\epsilon$.
The dependence on the squeezing phase $\phi$, however, is markedly branch-sensitive: Fig.~\ref{fig:S_vs_phi_EP} displays
an out-of-phase response between $\Delta=+\epsilon$ and $\Delta=-\epsilon$, where the phase values that maximize the
gauging eigenvalues on one branch tend to minimize them on the other. In other words, while the EP condition itself is
set purely by drift defectiveness, the magnitude and anisotropy of the diffusion gauge are strongly modulated by the bath
phase, and this modulation is exchanged between the two EP branches.

\subsection{Non-Markovian single-mode Gaussian channels: EPs and analytic noise gauging}
\label{subsec:NM_single_mode}

We illustrate the channel-level exceptional-point (EP) analysis and the associated noise-gauging covariance in the
simplest nontrivial setting: a single bosonic mode evolving under a \emph{family} of Gaussian channels
$\{\Psi_t\}_{t\ge 0}$ which is \emph{not assumed} to form a semigroup (hence non-Markovian in the channel sense). We use the
quadrature ordering $\hat x=(\hat q,\hat p)^{\T}$ with $\Sigma=\begin{pmatrix}0&1\\-1&0\end{pmatrix}$, and write each map as
\begin{equation}
\Psi_t \equiv (X_t,Y_t,\delta_t),\quad
X_t\in\mathbb R^{2\times 2},\ \ Y_t=Y_t^{\T}\in\mathbb R^{2\times 2},\ \ \delta_t\in\mathbb R^{2}.
\end{equation}
Its action on moments and complete positivity conditions are given by~\eqref{eq:GaussianChannelMoments} and \eqref{eq:CPconstraint}, respectively.
For one mode, any real $2\times 2$ matrix satisfies ${X\Sigma X^{\T}=(\det X)\Sigma}$, hence complete positivity  reduces to the scalar determinant condition (see~\cref{app:single_mode_CP_det})
\begin{equation}
Y_t\succeq 0,\quad \det Y_t \ge \Big(\frac{1-\det X_t}{2}\Big)^2.
\label{eq:CP_single_mode_det}
\end{equation}
For each fixed $t>0$, the channel EP structure (defectiveness and Jordan blocks of the superoperator) is governed by the
linear map $\xi\mapsto X_t^{\T}\xi$ acting on characteristic functions, and therefore by the eigen- and Jordan structure
of the drift matrix $X_t$ itself. Diffusion $Y_t$ dresses eigenoperators but, under the similarity gauging developed
above, does not shift EP locations. In particular, for a real $2\times2$ matrix, defectiveness occurs if and only if $X_t$
has a repeated eigenvalue and is not proportional to the identity.

Fix now a time $t>0$ and consider the \emph{single} Gaussian map $\Psi_t\equiv(X_t,Y_t,\delta_t)$. Assuming stability for
that single use,
\begin{equation}
\spr(X_t)<1,
\label{eq:stability_single_map}
\end{equation}
the Stein equation
\begin{equation}
S_t = X_t S_t X_t^{\T} + Y_t
\label{eq:Stein_single_mode_general}
\end{equation}
admits a unique real symmetric solution $S_t=S_t^{\T}$, given by~\eqref{eq:SteinSeries_general_app}, and satisfying
$S_t\succeq 0$ whenever $Y_t\succeq 0$.
Since $S_t$ has only three independent entries, the same solution can equivalently be obtained by solving an explicit
$3\times 3$ linear system, which is worked out in detail in~\cref{app:Stein_one_mode_vec}.

A particularly transparent simplification occurs directly on a defective (Jordan) drift. If, at a given parameter value,
$X_t$ is defective and admits a Jordan decomposition
\begin{equation}
X_t=\alpha_t\,(I_2+tN),\quad N\neq 0,\quad N^2=0,
\label{eq:X_Jordan_single_mode}
\end{equation}
one has ${(I_2+tN)^n=I_2+n t N}$ and therefore
\begin{equation}
X_t^n=\alpha_t^{\,n}\big(I_2+n t N\big).
\end{equation}
Substituting into \eqref{eq:SteinSeries_general_app}, namely
\begin{equation}
S_t=\sum_{n=0}^{\infty} X_t^{n}Y_t\big(X_t^{\T}\big)^{n},
\end{equation}
and using the standard sums with $\rho=\alpha_t^2$ (for ${|\alpha_t|<1}$, hence stability), 
${\sum_{n=0}^{\infty}\rho^n=\frac{1}{1-\rho}}$,
${\sum_{n=0}^{\infty} n\rho^n=\frac{\rho}{(1-\rho)^2}}$,
${\sum_{n=0}^{\infty} n^2\rho^n=\frac{\rho(1+\rho)}{(1-\rho)^3}}$
 yields the closed form
\begin{align}
&S_t
=
\frac{1}{1-\rho}\,Y_t
+\frac{\rho}{(1-\rho)^2}\,t\,(N Y_t + Y_t N^{\T}) \nonumber\\
& +\frac{\rho(1+\rho)}{(1-\rho)^3}\,t^2\,N Y_t N^{\T},
\label{eq:S_closed_on_Jordan}
\end{align}
where ${\rho=\alpha_t^2}$.
Equation~\eqref{eq:S_closed_on_Jordan} provides an analytic gauge covariance directly on the EP set.

\subsubsection{A concrete non-semigroup family: $X_t=\kappa(t)e^{tB(\lambda,\omega)}$}
\label{subsubsec:instantiation_expB_kappa}

We now instantiate this discussion with a two-parameter drift generator
\begin{equation}
B(\lambda,\omega):=
\begin{pmatrix}
\lambda & \omega\\
-\omega & -\lambda
\end{pmatrix},
\quad \Tr B = 0,
\label{eq:B_lambda_omega}
\end{equation}
and define a (generally non-semigroup) family
\begin{equation}
\Psi_t \equiv (X_t,Y_t,0),
\quad
X_t := \kappa(t)\,e^{tB(\lambda,\omega)}.
\label{eq:X_kappa_expB}
\end{equation}
Non-Markovianity is encoded by taking a \emph{memory factor} $\kappa(t)$ that is not exponential, so that in general
${\Psi_{t+s}\neq\Psi_t\circ\Psi_s}$.
A simple explicit choice is
\begin{equation}
\kappa(t)=\exp\!\big[-\gamma t + r\sin(\nu t)\big],
\quad 0<r<\gamma/\nu,
\label{eq:kappa_choice}
\end{equation}
which may be viewed as a phenomenological modulation capturing memory/feedback.

Since $B^2=(\lambda^2-\omega^2)I_2$, $e^{tB}$ is explicit. Defining $\nu:=\sqrt{\lambda^2-\omega^2}$, for $\lambda^2>\omega^2$
one has
\begin{equation}
e^{tB}
=
\cosh(\nu t)\,I_2 + \frac{\sinh(\nu t)}{\nu}\,B,
\label{eq:expB_hyp}
\end{equation}
while for $\lambda^2<\omega^2$ one sets $\mu:=\sqrt{\omega^2-\lambda^2}$ and replaces $\cosh(\nu t)\mapsto \cos(\mu t)$ and
$\sinh(\nu t)/\nu\mapsto \sin(\mu t)/\mu$. On the critical line $\lambda^2=\omega^2$ one has $B^2=0$ and hence
\begin{equation}
e^{tB}=I_2+tB,
\quad
X_t=\kappa(t)(I_2+tB).
\label{eq:EP_explicit_Xt}
\end{equation}
For any fixed $t>0$, $X_t$ then has a double eigenvalue $\kappa(t)$ and is not proportional to the identity whenever
$B\neq 0$, so it is defective with a size-2 Jordan block. Therefore,
\begin{equation}
\text{Channel EPs occur for }t>0\ \text{iff}\ \lambda^2=\omega^2\ \ (\lambda=\pm\omega),
\label{eq:EP_manifold_lines}
\end{equation}
excluding the trivial point $\omega=0$ where $B=0$.

\subsubsection{Diffusion models and analytic gauge covariance on the EP lines}
\label{subsubsec:diffusion_cases_reorg}

To quantify how diffusion enters without shifting the exceptional-point (EP) set, we consider three noise structures
$Y_t$ and the corresponding analytic expressions for the gauging covariance $S_t$, which measures the Gaussian smoothing
required to gauge away $Y_t$ at fixed time $t$.

For a single mode, complete positivity is equivalent to the determinant inequality \eqref{eq:CP_single_mode_det}.
In the numerical illustrations we choose diffusion strengths that are \emph{near minimal}, i.e.\ we take $Y_t$ to
essentially saturate the CP bound. To avoid working exactly on the CP boundary---where degeneracies can occur and the
Stein problem may become numerically ill-conditioned---we introduce a small buffer $\varepsilon>0$, ensuring that
$\det Y_t$ lies strictly above its minimal CP value.

\paragraph{(i) Isotropic diffusion.}
Take
\begin{equation}
Y_t = y(t)\,I_2,\quad y(t)\ge 0.
\label{eq:isotropicY_reorg}
\end{equation}
Since $\det(e^{tB})=1$, we have $\det X_t=\kappa(t)^2$ and \eqref{eq:CP_single_mode_det} reduces to
\begin{equation}
y(t)\ \ge\ \frac12|1-\det X_t|
\ =\ \frac12|1-\kappa(t)^2|.
\label{eq:CP_isotropic_reorg}
\end{equation}
In the numerical plots we choose the near-minimal strength
\begin{equation}
y(t)=\frac12|1-\kappa(t)^2|+\varepsilon,
\label{eq:y_isotropic_near_min}
\end{equation}
so that \eqref{eq:CP_isotropic_reorg} is satisfied with a strict margin.

On the EP manifold $\lambda^2=\omega^2$ one has $B^2=0$ and $X_t=\kappa(t)(I_2+tB)$, i.e.\ the Jordan form
\eqref{eq:X_Jordan_single_mode} with $\alpha_t=\kappa(t)$ and $N=B$. Substituting into \eqref{eq:S_closed_on_Jordan}
yields
\begin{equation}
S_t
=
\frac{y(t)}{1-\rho}\,I_2
+\frac{y(t)\,\rho}{(1-\rho)^2}\,t\,(B+B^{\T})
+\frac{y(t)\,\rho(1+\rho)}{(1-\rho)^3}\,t^2\,B B^{\T},
\label{eq:S_isotropic_EP_reorg}
\end{equation}
where $\rho=\kappa(t)^2$, giving $\lambda_{\min/\max}(S_t)$ analytically along the EP lines.

\paragraph{(ii) Anisotropic diffusion.}
Allow a general anisotropic diffusion matrix
\begin{equation}
Y_t=Y_t^{\T}\succeq 0,\quad\text{e.g. } Y_t=\mathrm{diag}\big(y_q(t),y_p(t)\big).
\label{eq:anisotropicY_reorg}
\end{equation}
Complete positivity is again equivalent to \eqref{eq:CP_single_mode_det}. In the numerical illustrations we fix
\begin{equation}
\det Y_t=\Big(\frac{1-\kappa(t)^2}{2}+\varepsilon\Big)^2,
\label{eq:det_target_aniso}
\end{equation}
and realize this determinant with a diagonal anisotropic choice (e.g.\ $y_q=g\,e^{s}$, $y_p=g\,e^{-s}$, so that
$\det Y_t=g^2$). On the EP manifold, the Jordan closed form \eqref{eq:S_closed_on_Jordan} applies with
$\alpha_t=\kappa(t)$ and $N=B$, giving
\begin{equation}
S_t
=
\frac{1}{1-\rho}\,Y_t
+\frac{\rho}{(1-\rho)^2}\,t\,(B Y_t + Y_t B^{\T})
+\frac{\rho(1+\rho)}{(1-\rho)^3}\,t^2\,B Y_t B^{\T},
\label{eq:S_anisotropic_EP_reorg}
\end{equation}
with $\rho=\kappa(t)^2$.

\paragraph{(iii) Drift-aligned structured diffusion.}
To emphasize branch sensitivity along $\lambda=\pm\omega$, we choose $Y_t$ to be \emph{built} from the drift tensor,
\begin{equation}
Y_t(\lambda,\omega)
=
\nu(\lambda,\omega)\Bigg(
I_2+\alpha\,\frac{B(\lambda,\omega)B(\lambda,\omega)^{\T}}
{\Tr\!\big[B(\lambda,\omega)B(\lambda,\omega)^{\T}\big]}
\Bigg),
\quad \alpha>0,
\label{eq:drift_alignedY_reorg}
\end{equation}
and fix $\nu(\lambda,\omega)$ by the near-minimal determinant target
\begin{equation}
\det Y_t(\lambda,\omega)=\Big(\frac{1-\kappa(t)^2}{2}+\varepsilon\Big)^2,
\label{eq:det_target_reorg}
\end{equation}
which places the model strictly inside the CP region.

On $\lambda=\pm\omega$, the off-diagonal correlations encoded in $B B^{\T}$ change sign between the two branches, and so do
the corresponding correlations in $Y_t$; consequently, the gauging covariance $S_t$ obtained from
\eqref{eq:S_anisotropic_EP_reorg} inherits a clear branch dependence.

\begin{figure}[!htbp]
  \centering

  \begin{subfigure}[t]{0.9\linewidth}
    \centering
    \includegraphics[width=\linewidth]{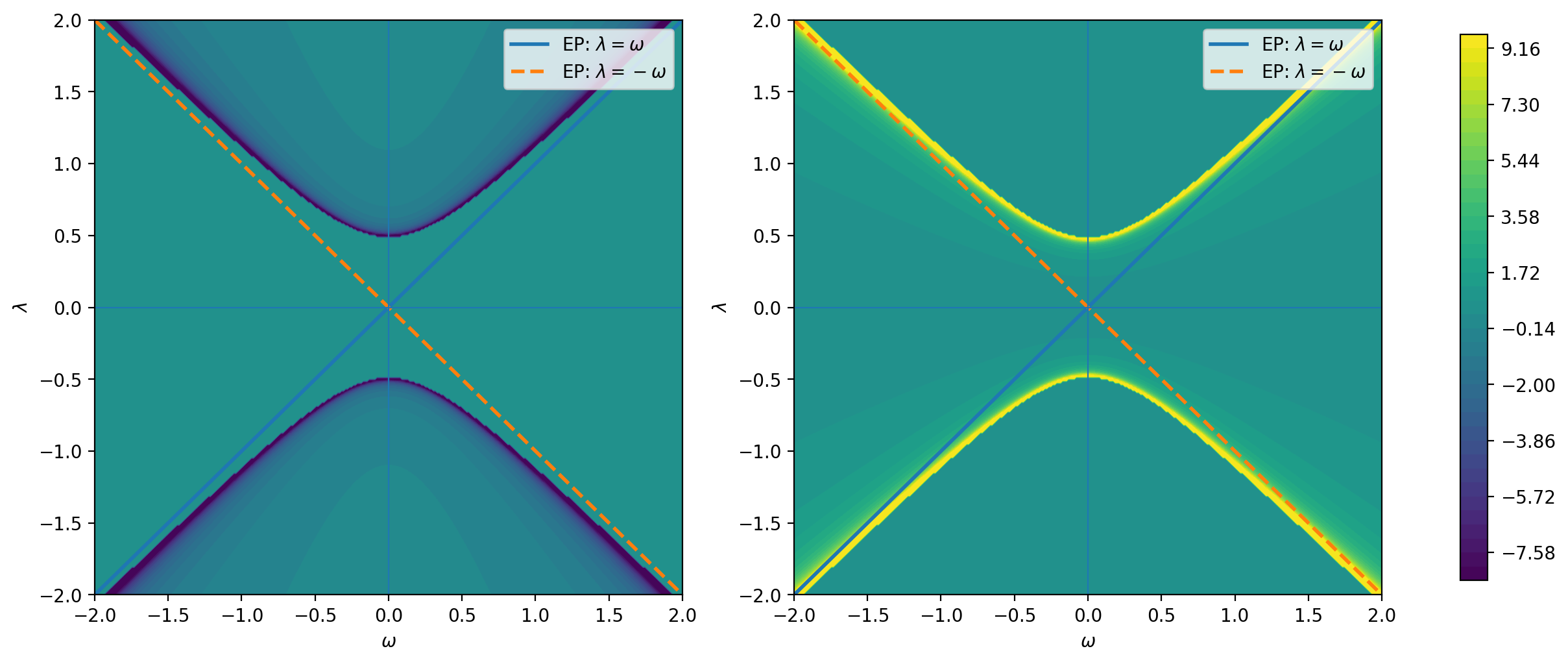}
    \caption{Isotropic diffusion: $Y_t=y(t)I_2$, with $y(t)$ saturating the CP bound \eqref{eq:CP_isotropic_reorg}.}
    \label{fig:St_eigs_iso}
  \end{subfigure}

  \vspace{0.8em}

  \begin{subfigure}[t]{0.9\linewidth}
    \centering
    \includegraphics[width=\linewidth]{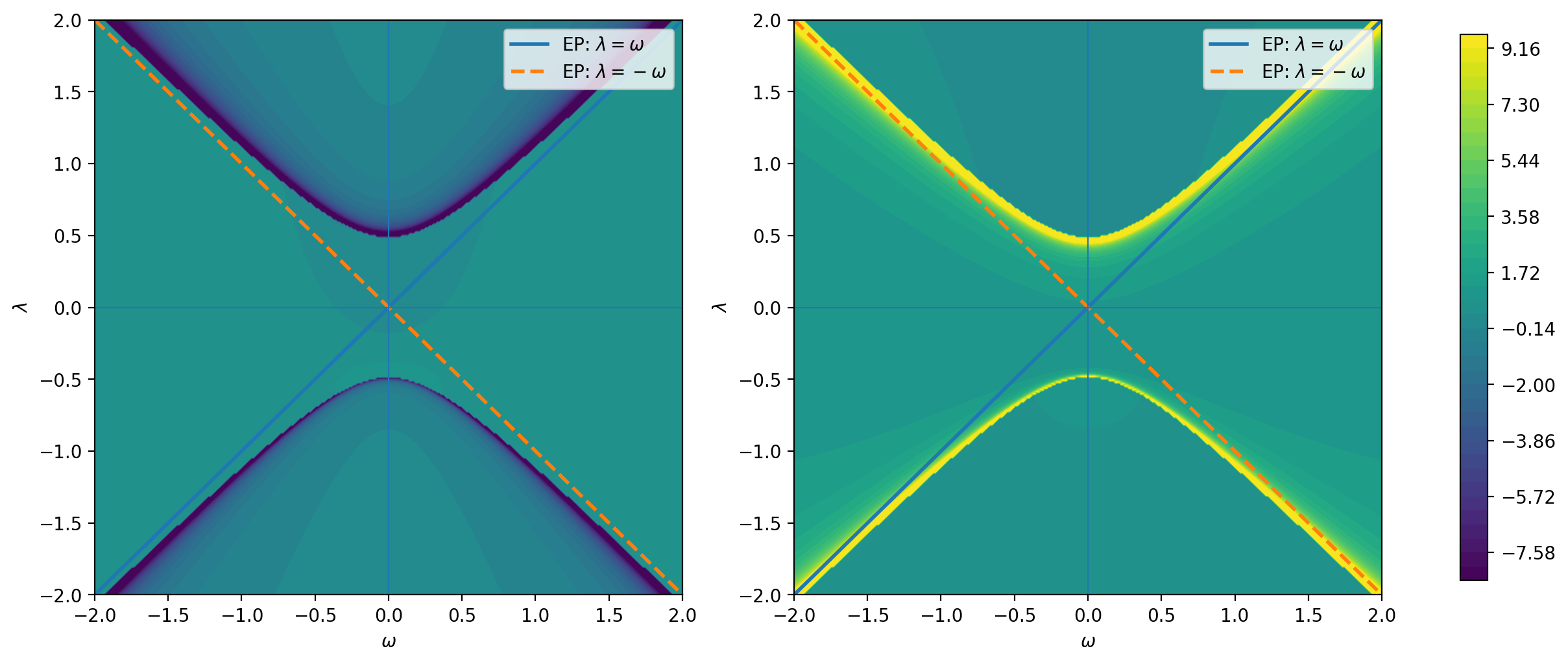}
    \caption{Anisotropic diffusion: $Y_t=Y_t^{\T}\succeq 0$, e.g.\ $Y_t=\diag\!\big(y_q(t),y_p(t)\big)$, constrained by
    \eqref{eq:CP_single_mode_det}.}
    \label{fig:St_eigs_aniso}
  \end{subfigure}

  \vspace{0.8em}

  \begin{subfigure}[t]{0.9\linewidth}
    \centering
    \includegraphics[width=\linewidth]{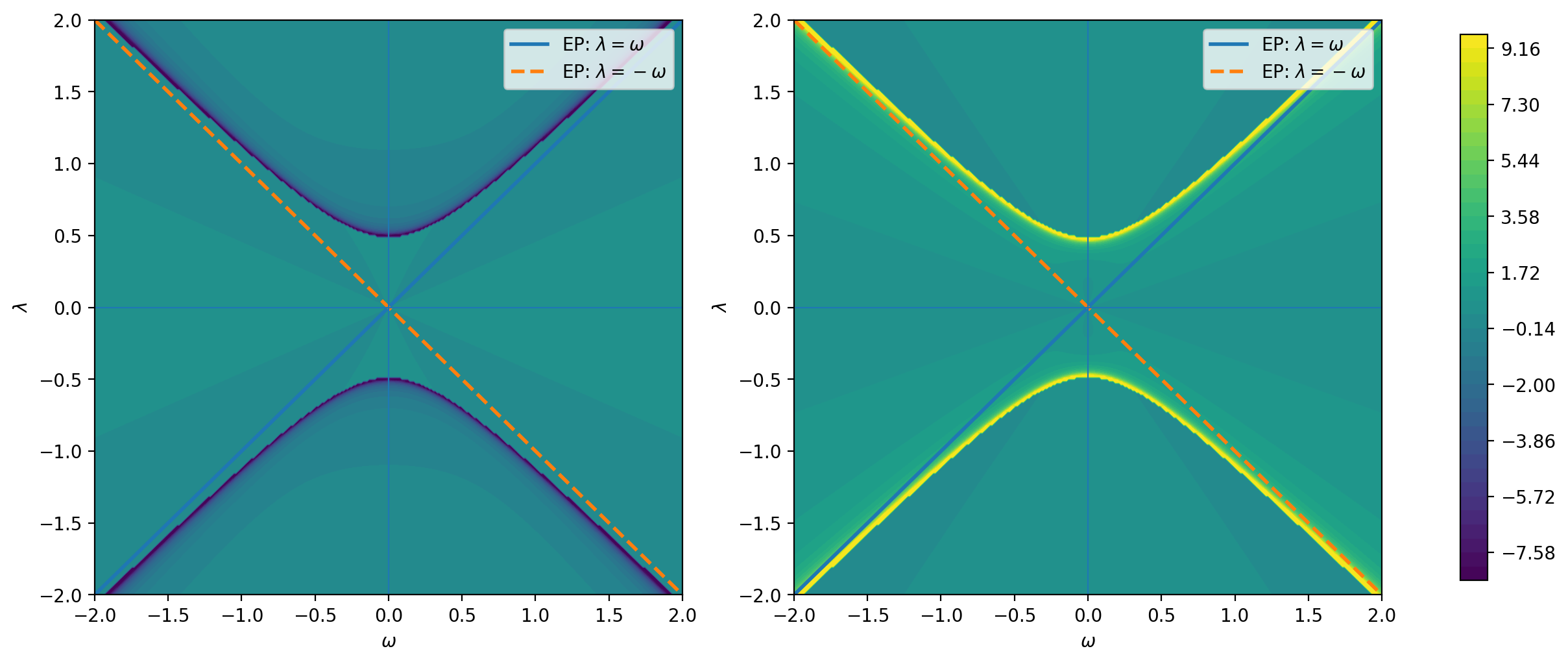}
    \caption{Drift-aligned structured diffusion: $Y_t(\lambda,\omega)$ from \eqref{eq:drift_alignedY_reorg} with
    $\nu(\lambda,\omega)$ fixed by \eqref{eq:det_target_reorg}.}
    \label{fig:St_eigs_driftaligned}
  \end{subfigure}

  \caption{\justifying Eigenvalues of the gauging covariance $S_t$ for a non-Markovian single-mode Gaussian channel.
  At fixed time $t$, the map is specified by $(X_t,Y_t)$ and $S_t$ is the unique real symmetric solution of the Stein
  equation $S_t=X_t S_t X_t^{\T}+Y_t$ in the stable regime $\spr(X_t)<1$.
  The plotted surfaces are obtained from the \emph{analytic} solution of this Stein problem (closed-form expressions in
  \cref{app:Stein_one_mode_vec}). In each panel, the left (right) subplot shows $\lambda_{\min}(S_t)$
  ($\lambda_{\max}(S_t)$) over the drift plane $(\lambda,\omega)$, with the EP branches $\lambda=\pm\omega$ overlaid
  (solid: $\lambda=\omega$, dashed: $\lambda=-\omega$). On the EP branches $X_t$ is defective and the Jordan closed form
  \eqref{eq:S_closed_on_Jordan} applies. Panels: (a) isotropic diffusion, (b) anisotropic diffusion, (c) drift-aligned
  structured diffusion.}
  \label{fig:St_eigs_three_diffusions}
\end{figure}

\begin{figure}[!htbp]
\centering

\begin{subfigure}[t]{0.9\columnwidth}
\centering
\includegraphics[width=\linewidth]{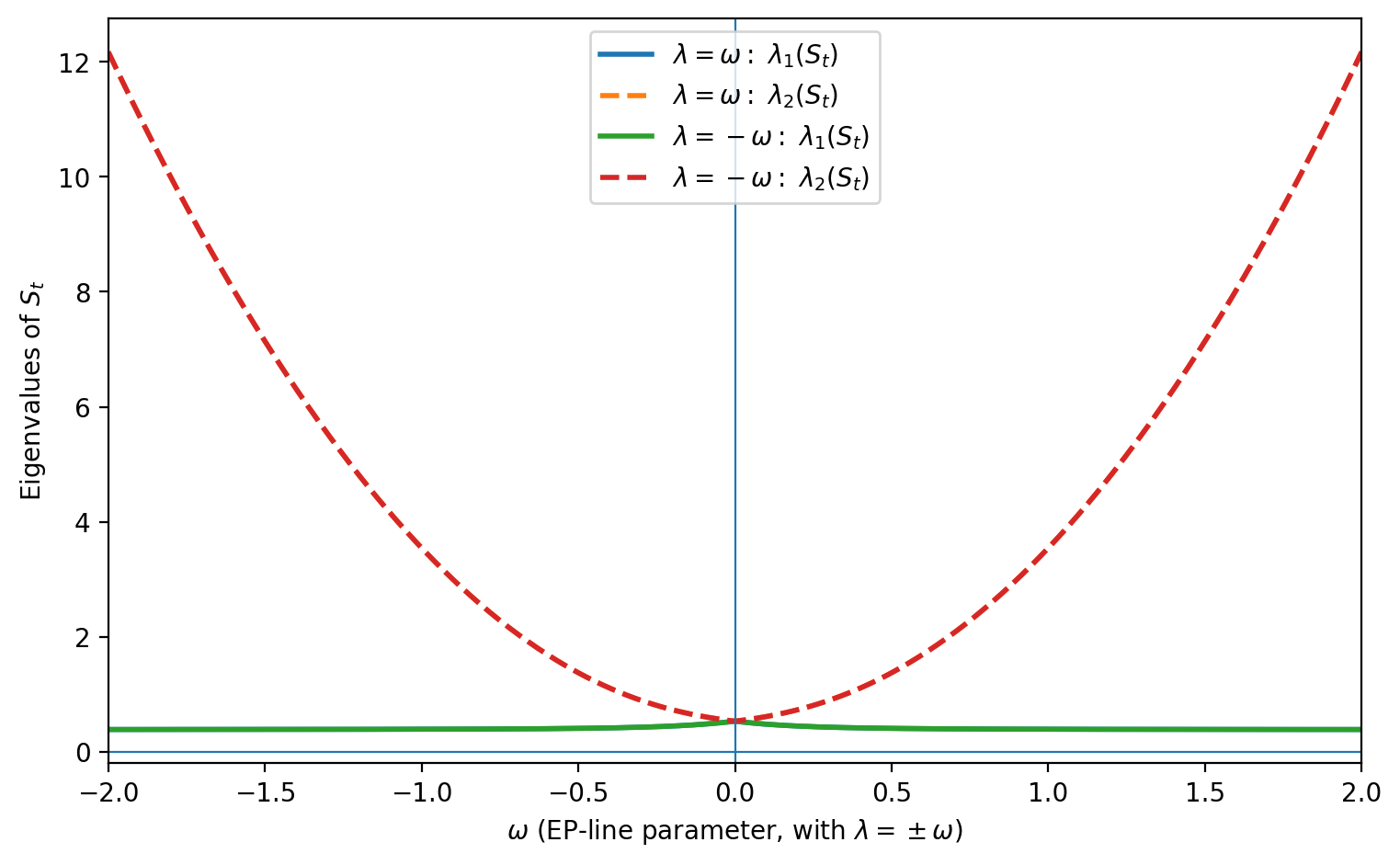}
\caption{\justifying Gauge eigenvalues $\lambda_{1,2}(S_t)$ on the EP branches $\lambda=\pm\omega$ for isotropic diffusion
$Y_t=y(t)I_2$, using the analytic Jordan-drift solution \eqref{eq:S_isotropic_EP_reorg}.}
\label{fig:S_eigs_iso_reorg}
\end{subfigure}

\vspace{0.8em}

\begin{subfigure}[t]{0.9\columnwidth}
\centering
\includegraphics[width=\linewidth]{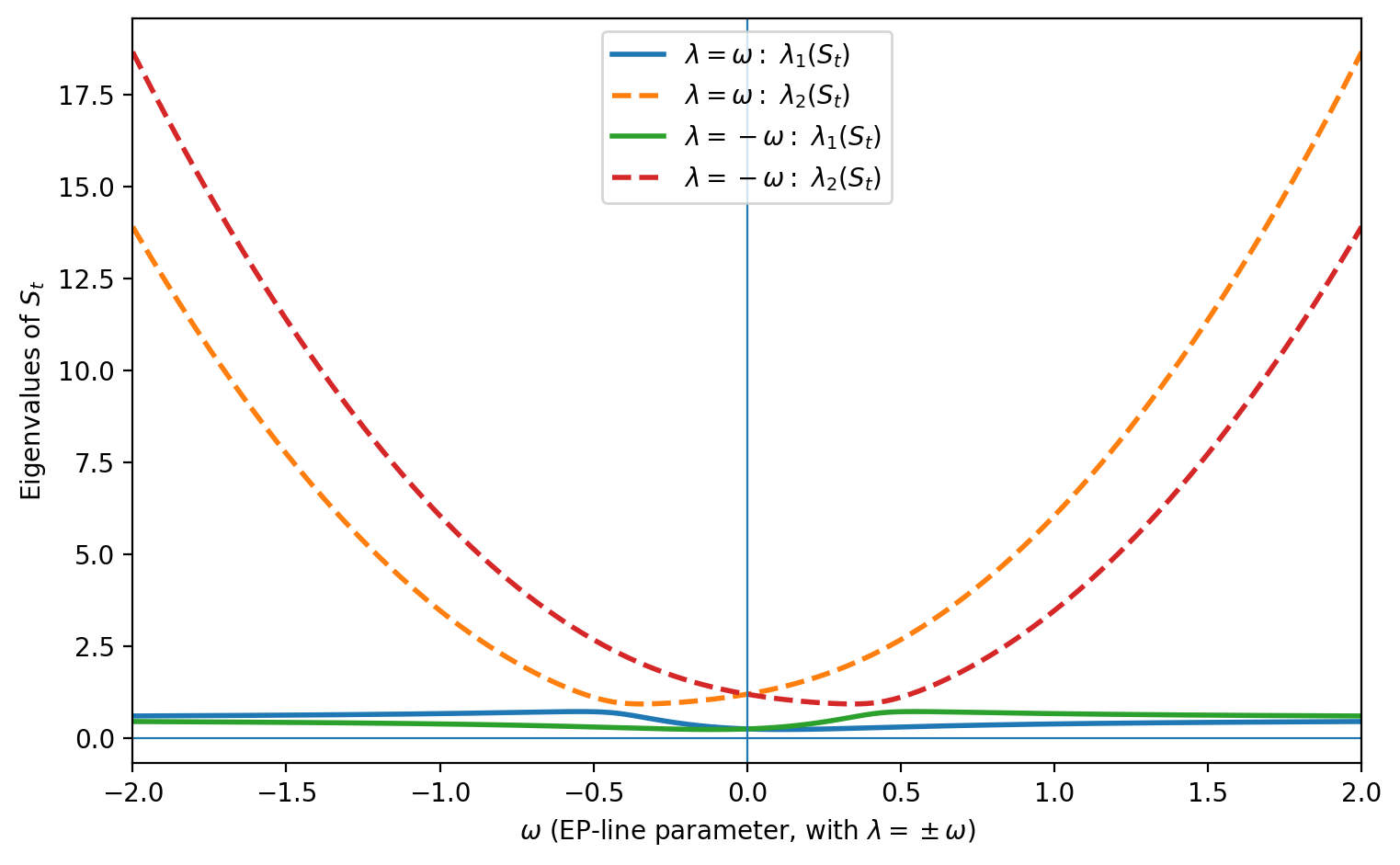}
\caption{\justifying Gauge eigenvalues $\lambda_{1,2}(S_t)$ on the EP branches $\lambda=\pm\omega$ for anisotropic diffusion
(e.g.\ $Y_t=\diag(y_q(t),y_p(t))$), from \eqref{eq:S_anisotropic_EP_reorg}.}
\label{fig:S_eigs_aniso_reorg}
\end{subfigure}

\vspace{0.8em}

\begin{subfigure}[t]{0.9\columnwidth}
\centering
\includegraphics[width=\linewidth]{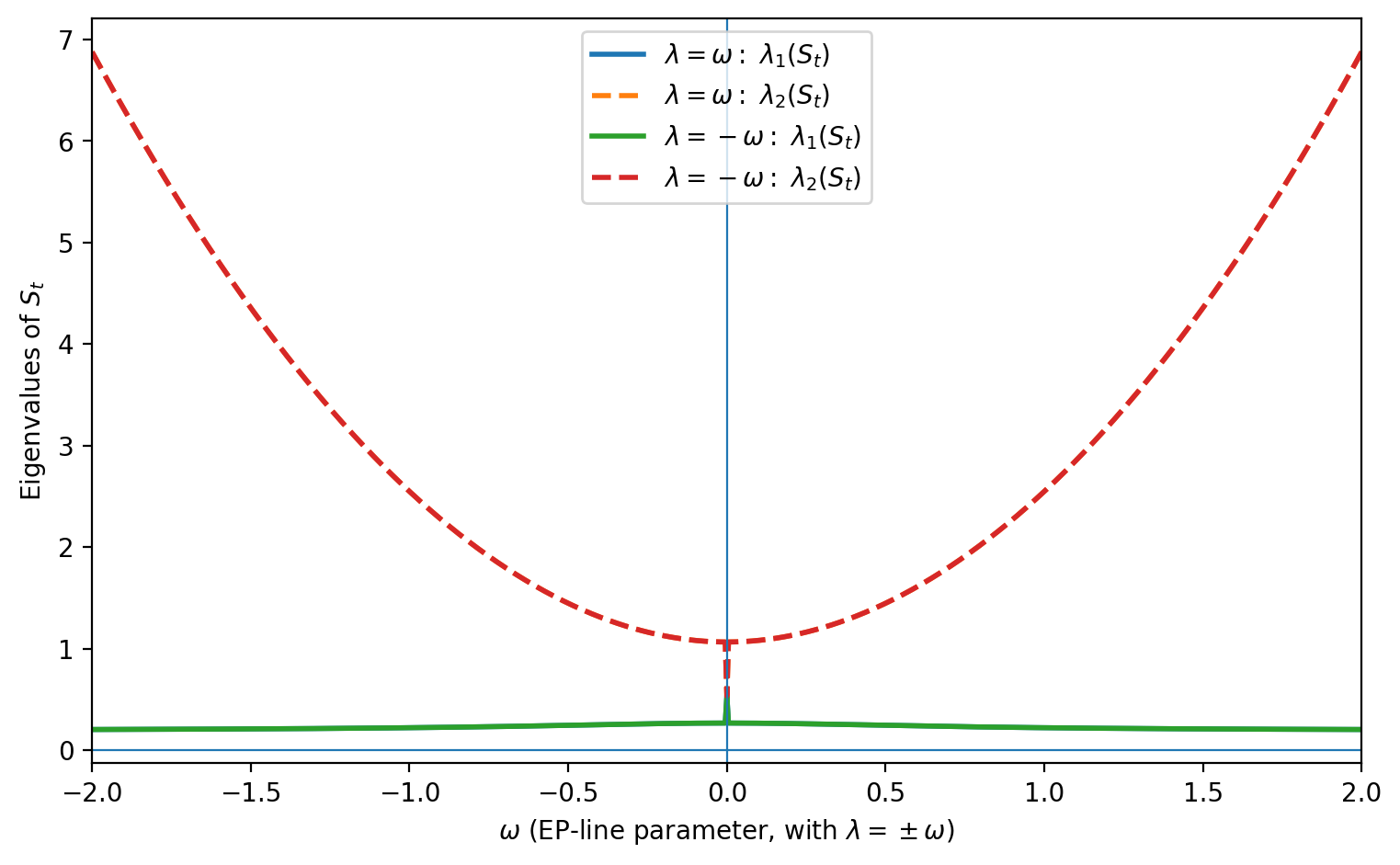}
\caption{\justifying Gauge eigenvalues $\lambda_{1,2}(S_t)$ on the EP branches $\lambda=\pm\omega$ for drift-aligned structured
diffusion \eqref{eq:drift_alignedY_reorg}, with $\nu(\lambda,\omega)$ fixed by the determinant target
\eqref{eq:det_target_reorg}.}
\label{fig:S_eigs_drift_aligned_reorg}
\end{subfigure}

\caption{\justifying Non-Markovian single-mode Gaussian channel at fixed time $t$: eigenvalues $\lambda_{1,2}(S_t)$ of the
gauging covariance $S_t$ along the exceptional-point branches $\lambda=\pm\omega$. The curves are obtained from the
\emph{analytic} solution of the Stein equation on a Jordan (defective) drift, with the corresponding closed-form
expressions summarized in \cref{app:Stein_one_mode_vec}. Panels: (a) isotropic diffusion $Y_t=y(t)I_2$; (b) anisotropic
diffusion $Y_t=\diag(y_q(t),y_p(t))$; (c) drift-aligned structured diffusion \eqref{eq:drift_alignedY_reorg}. The
branch-to-branch splitting in (c) reflects the sign change of the off-diagonal correlations built into $Y_t$ between the
two EP branches.}
\label{fig:S_EP_line_triptych_reorg}
\end{figure}

Relative to \cref{fig:St_eigs_three_diffusions}, where the three diffusion models yield broadly similar eigenvalue
landscapes over the full drift plane, restricting to the EP set $\lambda=\pm\omega$ reveals sharper
model-dependent differences.
In \cref{fig:S_eigs_iso_reorg} the two EP branches are indistinguishable: for isotropic diffusion $Y_t=y(t)I_2$, the
Jordan closed form \eqref{eq:S_isotropic_EP_reorg} depends on $B$ only through the symmetric combinations $B+B^{\T}$ and
$BB^{\T}$, and the gauge eigenvalues $\lambda_{1,2}(S_t)$ cannot resolve the sign change between $\lambda=\omega$ and
$\lambda=-\omega$.

In \cref{fig:S_eigs_aniso_reorg}, anisotropic diffusion lifts this branch insensitivity. While the two branches remain
qualitatively similar, their eigenvalue curves are displaced (most visibly near the origin), consistent with the
explicit mixed term $BY_t+Y_tB^{\T}$ in \eqref{eq:S_anisotropic_EP_reorg}.

Finally, \cref{fig:S_eigs_drift_aligned_reorg} isolates the effect of drift-locked correlations. By construction, the
structured diffusion \eqref{eq:drift_alignedY_reorg} is aligned with $BB^{\T}$ and inherits the EP geometry of the drift
tensor (in particular near $B\to 0$). As a result, the gauge eigenvalues display the strongest EP-controlled features,
reflecting that $Y_t$ is built from $B(\lambda,\omega)$ rather than imposed externally.

\subsubsection{When EPs do \emph{not} occur in common Gaussian channels}
\label{subsubsec:noEP_vs_EP_reorg}

The drift-controlled nature of defectiveness also clarifies why several standard one-mode channels often called
``damping'' or ``dephasing'' do not exhibit EPs. A phase-insensitive thermal-loss (attenuator) channel with transmissivity
$\eta\in[0,1]$ and bath occupancy $\bar n$ has
\begin{equation}
X=\sqrt{\eta}\,I_2,\quad
Y=\frac{1-\eta}{2}(2\bar n+1)\,I_2,\quad
\delta=0,
\end{equation}
and since $X\propto I_2$ for all $\eta$ it is always diagonalizable; hence no EP manifold occurs in this family. Likewise,
a pure quadrature-diffusion channel of the form
\begin{equation}
X=I_2,\quad
Y=\begin{pmatrix}0&0\\0&\sigma^2\end{pmatrix},\quad
\delta=0
\end{equation}
is EP-free because $X=I_2$ remains diagonalizable and defectiveness cannot arise; here $Y$ broadens the state without
generating non-normal drift.

To obtain EPs one needs a drift that can become non-diagonalizable. A canonical continuous-time example is the critically
damped oscillator, for which the drift generator becomes defective at critical damping; equivalently $X_t=e^{At}$ acquires
a Jordan form $X_t=e^{-\omega_0 t}(I_2+tN)$ with $N^2=0$, producing EP physics (Jordan blocks and polynomially dressed
transients). Diffusion does not shift the EP condition but determines the gauging covariance via the corresponding Stein
equation at fixed $t$, or Lyapunov equation in semigroup settings.

\section{Conclusions and outlook}
\label{sec:conclusions}

We introduced a channel-level framework for diagnosing exceptional points (EPs) in multimode bosonic Gaussian dynamics.
Our central message is a clean drift--diffusion separation of spectral data that can be formulated both at the generator
level (continuous time) and at the single-map level (discrete time).

In continuous time, we considered Gaussian Markov dynamics generated by the diffusion equations
\eqref{eq:momentODEs_abs}. 
In the Weyl (symmetrically ordered) characteristic-function representation the corresponding Liouvillian acts as
an Ornstein--Uhlenbeck operator \eqref{eq:OU_generator_general}. Under the stability condition that the drift matrix $A$ is
Hurwitz, the unique symmetric solution $S$ of the Lyapunov equation $AS+SA^{\T}+D=0$ defines an explicit Gaussian
similarity transform on characteristic functions, $\widetilde\chi=\exp(\tfrac12\xi^{\T}S\xi)\chi$, which gauges away the
quadratic diffusion term and produces the drift-only generator \eqref{eq:OU_gauged_generator}. Equivalently, at the
channel level this Lyapunov covariance defines a time-independent smoothing map $\mathcal V_S\equiv(I,S,0)$ such that the
entire semigroup is similar to a diffusion-free semigroup,
$\widetilde\Psi_t=\mathcal V_S^{-1}\circ\Psi_t\circ\mathcal V_S\equiv(\e^{At},0,\delta_t)$ for all $t\ge0$.
Since similarity preserves eigenvalues and Jordan structure, Liouvillian EP manifolds are drift-controlled: they coincide
with the defect set of $A$, while diffusion enters only through the dressing covariance $S$ (and hence through the steady
state and the structure of eigenoperators). This provides a phase-space and channel-composition formulation of the
noise-independence of spectral data emphasized from complementary viewpoints such as third quantization
\cite{McDonaldClerk2023}.

Motivated by this continuous-time structure, we then extended the same separation principle to discrete time at the level
of a single Gaussian channel. For any stable Gaussian channel $\Psi\equiv(X,Y,\delta)$ with $\spr(X)<1$, the unique
symmetric solution $S$ of the Stein equation $S=XSX^{\T}+Y$ defines a Gaussian ``smoothing'' map
$\mathcal V_S\equiv(I,S,0)$ such that
$\widetilde\Psi=\mathcal V_S^{-1}\circ\Psi\circ\mathcal V_S\equiv(X,0,\delta)$.
This is a genuinely single-map statement: it does not require $\Psi$ to be embeddable in a Markov semigroup, and is
therefore logically independent of the Lyapunov gauging theorem. Again, because the transformation is a similarity, EP
locations and their order are governed by defectiveness of $X$, while diffusion $Y$ affects only the steady state and the
dressing of eigenoperators through the gauging covariance $S$.

Our single-mode applications illustrated how the gauging covariance packages physically relevant noise information without
shifting EP conditions. In the Markovian squeezed-reservoir Lindbladian, the drift exhibits an EP on $\Delta=\pm\epsilon$,
independently of the squeezing parameters, while the Lyapunov gauge covariance $S$ is generically anisotropic and phase
sensitive, reflecting reservoir squeezing. In the non-Markovian setting, we showed how EP manifolds of the fixed-time map
$\Psi_t$ are obtained analytically from the defectiveness of $X_t$, and how the corresponding Stein solution $S_t$
quantifies the smoothing required to gauge away diffusion for that single channel use, including compact closed forms
directly on the EP set. Finally, the drift-control criterion makes transparent why many standard one-mode phase-insensitive
``damping'' and diffusion-only channels are EP-free (their drift is proportional to the identity), and highlights minimal
non-normal mechanisms---such as critical damping---where defective drift and Jordan-block transients arise naturally.

Several extensions are immediate. It would be valuable to relax the stability assumptions ($\spr(X)<1$ or
$\Ree\,\lambda(A)<0$) and understand how the gauging picture evolves in marginally stable or amplifying regimes, where the
relevant similarity transformations may become unbounded and EP physics interfaces with thresholds and gain. On the
multimode side, organizing EP manifolds by equivalence classes of drift maps (including symplectic structure and
mode-coalescence patterns) and then incorporating diffusion through the associated gauging covariance $S$ should enable a
systematic classification of EP scenarios for general Gaussian channels. On the experimental front, connecting these
channel-level diagnostics to measurable signatures---polynomial prefactors in transient moment dynamics, enhanced
parameter sensitivity near drift defectiveness, and robustness of EP locations under changes in bath temperature and
squeezing (which enter diffusion but not the drift-controlled EP manifold)---is a natural next step.

A further direction is to explore potential connections between EPs and phase-transition-like behavior in families of
Gaussian channels. Since EPs are non-analytic points of spectral data (coalescing eigenvalues/eigenvectors and changing
Jordan structure), it is natural to ask whether they can organize sharper crossovers or non-analyticities in operational
quantities built from channels---for example, in fixed-point covariances, mixing times, or long-time limits under repeated
channel use---especially in multimode settings or scaling limits where collective behavior becomes prominent.

\acknowledgments{We acknowledge financial support from the EU (ERC grant TIMELIGHT, GA 101115792) and from MCIUN/AEI (PID2022-141036NA-I00 through MCIUN/AEI/10.13039/501100011033 and FSE+; RYC2021-031568-I; Programme for Units of Excellence in R\&D CEX2023-001316-M). We also thank the members of the TIMELIGHT project at IFIMAC for their support and discussions, as well as Donato Farina, Vasco Cavina, and the Quantum Information group at Scuola Normale di Pisa for insightful conversations.}



\onecolumngrid
\appendix

\section{Why the Liouvillian spectrum and Jordan structure are drift-controlled}
\label{app:drift_controls_spectrum}

This Appendix spells out the points used in the main text after
Eq.~\eqref{eq:OU_gauged_generator}. The key idea is simple: by a change of
representation (a ``gauge'' transformation of the characteristic function) one
can remove the diffusion term from the generator. Since this change is
invertible, it does not alter eigenvalues or Jordan block sizes. As a result,
the spectrum and the appearance of exceptional points are governed entirely by
the drift matrix $A$.
More explicitly, we show: (i) the Ornstein--Uhlenbeck generator is related by an
invertible transformation to a first-order operator involving only the drift,
so eigenvalues and Jordan blocks are fixed by $A$; (ii) when $u=0$ and $A$ is
diagonalizable, the Liouvillian spectrum is obtained by additive combinations
of the eigenvalues of $A$, as in Eq.~\eqref{eq:Liouvillian_spec_from_A_general};
and (iii) if $A$ is not diagonalizable, then the Liouvillian is not
diagonalizable either (i.e., one has exceptional points), regardless of the
diffusion matrix $D$.

\begin{remark*}[Where the identities are meant to hold]
All operator manipulations below are meant on a convenient class of test
functions of $\xi$ (for instance, polynomials in $\xi$, or Schwartz functions).
On such functions, multiplying by $\exp(\tfrac12\xi^{\T}S\xi)$ is well defined
and invertible. This is enough for the conclusions we use in the main text:
an invertible change of representation preserves eigenvalues and Jordan-block
sizes.
\end{remark*}

\subsection{A change of representation removes diffusion and keeps the same Jordan blocks}
\label{app:similarity_preserves_Jordan}

Consider the Ornstein--Uhlenbeck evolution of the Weyl (symmetrically ordered)
characteristic function $\chi(\xi,t)$,
\begin{equation}
\partial_t \chi(\xi,t)
=
\mathcal L\,\chi(\xi,t)
=
-\tfrac12\,\xi^{\T}D\,\xi\;\chi(\xi,t)
+\big(A^{\T}\xi\big)\!\cdot\nabla_\xi \chi(\xi,t)
+\ii\,u^{\T}\xi\;\chi(\xi,t),
\label{eq:OU_generator_app}
\end{equation}
where $A\in\mathbb R^{2N\times 2N}$, $D=D^{\T}\in\mathbb R^{2N\times 2N}$, and
$u\in\mathbb R^{2N}$ are constant.

Assume $A$ is Hurwitz so that the Lyapunov equation
\begin{equation}
AS+SA^{\T}+D=0
\label{eq:Lyapunov_app}
\end{equation}
admits a unique real symmetric solution $S=S^{\T}$. Define the multiplication
operator
\begin{equation}
(Tf)(\xi):=\exp\!\big(\tfrac12\xi^{\T}S\xi\big)\,f(\xi),
\quad
T^{-1}f=\exp\!\big(-\tfrac12\xi^{\T}S\xi\big)\,f(\xi).
\label{eq:T_def_app}
\end{equation}

\begin{proposition}[Removing diffusion without changing eigenvalues or Jordan blocks]
\label{prop:similarity_preserves_Jordan}
With $T$ as in \eqref{eq:T_def_app}, the transformed generator is
\begin{equation}
\widetilde{\mathcal L}=T\,\mathcal L\,T^{-1}
=
(A^{\T}\xi)\cdot\nabla_\xi + \ii\,u^{\T}\xi.
\label{eq:similarity_L_app}
\end{equation}
In particular, $\mathcal L$ and $\widetilde{\mathcal L}$ have the same
eigenvalues (with the same multiplicities) and the same Jordan-block sizes.
Equivalently, the locations of exceptional points are identical for
$\mathcal L$ and $\widetilde{\mathcal L}$.
\end{proposition}

\begin{proof}
Let $g=T^{-1}f$, i.e.\ $f=e^{\frac12\xi^{\T}S\xi}g$. A direct derivative gives
\[
\nabla_\xi f
=e^{\frac12\xi^{\T}S\xi}\bigl(\nabla_\xi g+S\xi\,g\bigr).
\]
Substituting into \eqref{eq:OU_generator_app} yields
\begin{align*}
\mathcal L f
&=e^{\frac12\xi^{\T}S\xi}\Bigl[
-\tfrac12\,\xi^{\T}D\xi\,g
+(A^{\T}\xi)\cdot(\nabla_\xi g+S\xi\,g)
+\ii\,u^{\T}\xi\,g\Bigr]\\
&=e^{\frac12\xi^{\T}S\xi}\Bigl[
(A^{\T}\xi)\cdot\nabla_\xi g
+\Bigl(-\tfrac12\,\xi^{\T}D\xi+\xi^{\T}AS\xi\Bigr)g
+\ii\,u^{\T}\xi\,g\Bigr].
\end{align*}
Using $S=S^{\T}$, we may symmetrize $\xi^{\T}AS\xi$ as
$\xi^{\T}AS\xi=\tfrac12\xi^{\T}(AS+SA^{\T})\xi$. The Lyapunov identity
$AS+SA^{\T}=-D$ from \eqref{eq:Lyapunov_app} then cancels the quadratic term,
leaving \eqref{eq:similarity_L_app}.

Finally, since $T$ is invertible on the chosen test functions, it maps any
eigenvector (and any Jordan chain) of $\mathcal L$ to an eigenvector (and a
Jordan chain of the same length) of $\widetilde{\mathcal L}$, and conversely via
$T^{-1}$. Hence eigenvalues and Jordan-block sizes coincide.
\end{proof}

\subsection{Additive spectrum when $u=0$ and $A$ is diagonalizable}
\label{app:additive_spectrum}

Set $u=0$. Then the transformed generator reduces to
\begin{equation}
\widetilde{\mathcal L}=(A^{\T}\xi)\cdot\nabla_\xi.
\end{equation}
Assume $A$ is diagonalizable. Choose an invertible matrix $V$ such that
\begin{equation}
A^{\T}=V\Lambda V^{-1},
\quad
\Lambda=\mathrm{diag}(\lambda_1,\dots,\lambda_{2N}),
\end{equation}
and introduce coordinates $\eta:=V^{-1}\xi$. By the chain rule,
\begin{equation}
\nabla_\xi = V^{-{\T}}\nabla_\eta,
\quad
A^{\T}\xi = V\Lambda\eta,
\end{equation}
so
\begin{align}
\widetilde{\mathcal L}
&=(V\Lambda\eta)\cdot(V^{-{\T}}\nabla_\eta)
=(\Lambda\eta)\cdot\nabla_\eta
=\sum_{j=1}^{2N}\lambda_j\,\eta_j\,\partial_{\eta_j}.
\label{eq:Ltilde_Euler_app}
\end{align}

Consider monomials
\begin{equation}
m_{\mathbf n}(\eta):=\prod_{j=1}^{2N}\eta_j^{n_j},
\quad \mathbf n=(n_1,\dots,n_{2N})\in\mathbb N_0^{2N}.
\end{equation}
Since $\eta_j\partial_{\eta_j}$ simply counts powers of $\eta_j$, we obtain
\begin{equation}
\widetilde{\mathcal L}\,m_{\mathbf n}
=
\Big(\sum_{j=1}^{2N} n_j\,\lambda_j\Big)m_{\mathbf n}.
\end{equation}
Thus
\begin{equation}
\mathrm{spec}(\widetilde{\mathcal L})
=
\Big\{\;\sum_{j=1}^{2N} n_j\,\lambda_j(A)\ :\ n_j\in\mathbb N_0\;\Big\}.
\label{eq:spec_Ltilde_additive_app}
\end{equation}
By Proposition~\ref{prop:similarity_preserves_Jordan}, the same set is the
spectrum of $\mathcal L$, yielding Eq.~\eqref{eq:Liouvillian_spec_from_A_general}.

\paragraph*{Remark (the role of $u$).}
When $u\neq 0$, the term $\ii\,u^{\T}\xi$ mixes different polynomial degrees, so
monomials are no longer eigenfunctions. This does \emph{not} change the main
message of Proposition~\ref{prop:similarity_preserves_Jordan}: diffusion $D$
does not affect eigenvalues or Jordan-block sizes.

\subsection{If $A$ is not diagonalizable, neither is the Liouvillian}
\label{app:defective_A_implies_EP}

If $A$ is not diagonalizable, then the drift part already forces the Liouvillian
to be non-diagonalizable. To see the mechanism transparently, set $u=0$ and
focus on a Jordan block of $A^{\T}$:
\begin{equation}
A^{\T}=\lambda I+N,
\quad
N\neq 0,
\quad
N^{k}=0\ \text{for some }k\ge 2 .
\label{eq:Jordan_block_AT_main}
\end{equation}
Then
\begin{equation}
\widetilde{\mathcal L}=((\lambda I+N)\xi)\!\cdot\nabla_\xi
=\lambda\,(\xi\!\cdot\nabla_\xi)+(N\xi)\!\cdot\nabla_\xi.
\label{eq:Ltilde_split_main}
\end{equation}

Fix $\ell\in\mathbb N_0$ and let $\mathcal P_\ell$ be the finite-dimensional
space of homogeneous polynomials of total degree $\ell$ in the components of
$\xi$. On $\mathcal P_\ell$ the operator $\xi\!\cdot\nabla_\xi$ acts as
multiplication by $\ell$. The remaining term
\[
\mathcal N_\ell:=(N\xi)\!\cdot\nabla_\xi
\]
also maps $\mathcal P_\ell$ into itself and is nilpotent there (it shifts
monomials along a finite chain because $N$ is nilpotent). Therefore, on
$\mathcal P_\ell$ we have
\begin{equation}
\widetilde{\mathcal L}\big|_{\mathcal P_\ell}=\ell\lambda\,I+\mathcal N_\ell,
\quad \mathcal N_\ell\neq 0,\quad \mathcal N_\ell^{\,m}=0\ \text{for some }m,
\label{eq:Ltilde_restriction_main}
\end{equation}
which means $\widetilde{\mathcal L}$ has a nontrivial Jordan block at
eigenvalue $\ell\lambda$. By Proposition~\ref{prop:similarity_preserves_Jordan},
the same conclusion holds for $\mathcal L$: if $A$ is not diagonalizable, then
the Liouvillian has exceptional points, independently of the diffusion matrix
$D$.

For intuition, in the simplest $2\times2$ case one may take
$A^{\T}=\lambda I+N$ with
$N=\begin{pmatrix}0&1\\0&0\end{pmatrix}$. Writing $\xi=(\xi_1,\xi_2)$,
Eq.~\eqref{eq:Ltilde_split_main} becomes
\begin{equation}
\widetilde{\mathcal L}=\lambda(\xi_1\partial_{\xi_1}+\xi_2\partial_{\xi_2})+\xi_2\,\partial_{\xi_1}.
\label{eq:Ltilde_2x2_explicit_main}
\end{equation}
On $\mathcal P_\ell$, a convenient basis is
$p_j(\xi)=\xi_1^{\ell-j}\xi_2^{j}$ for $j=0,\dots,\ell$, and one finds
\begin{equation}
(\widetilde{\mathcal L}-\ell\lambda)\,p_j=(\ell-j)\,p_{j+1},
\quad j=0,\dots,\ell-1,
\quad (\widetilde{\mathcal L}-\ell\lambda)\,p_\ell=0,
\label{eq:Jordan_chain_polynomials_main}
\end{equation}
so these polynomials form a Jordan chain of length $\ell+1$ at eigenvalue
$\ell\lambda$.

The coefficient pattern in \eqref{eq:Jordan_chain_polynomials_main} is the standard action of a $2\times2$ Jordan (nilpotent)
block on the $\ell$th symmetric power: on the homogeneous subspace of degree $\ell$ with basis
$v_j:=\xi_1^{\ell-j}\xi_2^{j}$ one has $(\xi_2\partial_{\xi_1})v_j=(\ell-j)\,v_{j+1}$, producing a single Jordan chain of
length $\ell+1$. This is written explicitly in the classical polynomial model of the $\mathfrak{sl}_2$-representation on
$k[\xi_1,\xi_2]$; see, e.g.,Ref.~\cite{BurdeLieAlgNotes}, Sec.~1.5, where for $v_j=Y^{j}X^{m-j}$ one finds
$y\cdot v_j=(m-j)\,v_{j+1}$.

\subsection{Exceptional points of the \emph{Liouvillian} for the squeezed-reservoir model}
\label{app:Liouvillian_EPs_squeezed}

For the squeezed-reservoir master equation~\eqref{eq:ME_squeezed_res} the
characteristic function obeys the Ornstein--Uhlenbeck equation
\begin{equation}
\partial_t \chi(\xi,t)
=
-\tfrac12\,\xi^{\T}D\,\xi\;\chi(\xi,t)
+\big(A^{\T}\xi\big)\!\cdot\nabla_\xi \chi(\xi,t)
+\ii\,u^{\T}\xi\;\chi(\xi,t),
\label{eq:OU_squeezed_app}
\end{equation}
with $u=0$ in the present example. In the stable regime $A$ is Hurwitz, so the
Lyapunov equation $AS+SA^{\T}+D=0$ admits a unique symmetric solution $S=S^{\T}$.
Defining the transformed characteristic function
\begin{equation}
\widetilde\chi(\xi,t):=\exp\!\big(\tfrac12\xi^{\T}S\xi\big)\,\chi(\xi,t),
\end{equation}
one finds
\begin{equation}
\partial_t \widetilde\chi(\xi,t)=\big(A^{\T}\xi\big)\!\cdot\nabla_\xi \widetilde\chi(\xi,t),
\end{equation}
so $\mathcal L$ is related by an invertible change of representation to
$\widetilde{\mathcal L}=(A^{\T}\xi)\cdot\nabla_\xi$, and therefore has the same
eigenvalues and Jordan blocks.

For the drift matrix~\eqref{eq:A_DPA_squeezed},
\begin{equation}
\lambda_\pm(A)= -\frac{\kappa}{2}\pm \sqrt{\epsilon^{2}-\Delta^{2}}.
\end{equation}
Hence the Liouvillian is non-diagonalizable exactly when $A$ is
non-diagonalizable, namely on
\begin{equation}
\Delta^2=\epsilon^2,\quad \epsilon\neq 0,
\end{equation}
independently of the diffusion parameters $(r,\phi)$ (which only enter through
$S$).

\section{Gaussian composition rule}
\label{app:composition_rule}

\begin{proposition}[Composition of multimode Gaussian channels]
\label{prop:Gaussian_composition_rule}
Let $\Psi_1\equiv (X_1,Y_1,\delta_1)$ and $\Psi_2\equiv (X_2,Y_2,\delta_2)$ be Gaussian channels acting on Weyl
(symmetrically ordered) characteristic functions as
\begin{equation}
\chi_{\Psi(\rho)}(\xi)
=
\exp\!\Big(-\tfrac12\,\xi^{\T}Y\,\xi + \ii\,\delta^{\T}\xi\Big)\;
\chi_\rho(X^{\T}\xi),
\quad \xi\in\mathbb R^{2N}.
\label{eq:GaussianChannelChi_app}
\end{equation}
Then the composition $\Psi_{21}:=\Psi_2\circ\Psi_1$ is Gaussian with parameters
\begin{equation}
(X_{21},Y_{21},\delta_{21})
=
\bigl(X_2X_1,\; X_2Y_1X_2^{\T}+Y_2,\; X_2\delta_1+\delta_2\bigr).
\label{eq:Gaussian_composition_rule_app}
\end{equation}
\end{proposition}

\begin{proof}
Let $\rho$ be an arbitrary trace-class operator and fix $\xi\in\mathbb R^{2N}$. By applying
\eqref{eq:GaussianChannelChi_app} to $\Psi_2$ we obtain
\begin{equation}
\chi_{\Psi_{21}(\rho)}(\xi)
=
\chi_{\Psi_2(\Psi_1(\rho))}(\xi)
=
\exp\!\Big(-\tfrac12\,\xi^{\T}Y_2\,\xi + \ii\,\delta_2^{\T}\xi\Big)\;
\chi_{\Psi_1(\rho)}(X_2^{\T}\xi).
\label{eq:comp_step1_app}
\end{equation}
Applying \eqref{eq:GaussianChannelChi_app} again, now to $\Psi_1$ evaluated at $X_2^{\T}\xi$, gives
\begin{align}
\chi_{\Psi_1(\rho)}(X_2^{\T}\xi)
&=
\exp\!\Big(-\tfrac12\,(X_2^{\T}\xi)^{\T}Y_1\,(X_2^{\T}\xi)
+ \ii\,\delta_1^{\T}(X_2^{\T}\xi)\Big)\;
\chi_\rho\!\big(X_1^{\T}X_2^{\T}\xi\big)\nonumber\\
&=
\exp\!\Big(-\tfrac12\,\xi^{\T}(X_2Y_1X_2^{\T})\,\xi
+ \ii\,(X_2\delta_1)^{\T}\xi\Big)\;
\chi_\rho\!\big((X_2X_1)^{\T}\xi\big).
\label{eq:comp_step2_app}
\end{align}
Substituting \eqref{eq:comp_step2_app} into \eqref{eq:comp_step1_app} and collecting the exponential prefactors yields
\begin{equation}
\chi_{\Psi_{21}(\rho)}(\xi)
=
\exp\!\Big(-\tfrac12\,\xi^{\T}\big(Y_2+X_2Y_1X_2^{\T}\big)\xi
+ \ii\,(\delta_2+X_2\delta_1)^{\T}\xi\Big)\;
\chi_\rho\!\big((X_2X_1)^{\T}\xi\big).
\end{equation}
Comparing with the defining form \eqref{eq:GaussianChannelChi_app} identifies
\[
X_{21}=X_2X_1,\quad
Y_{21}=X_2Y_1X_2^{\T}+Y_2,\quad
\delta_{21}=X_2\delta_1+\delta_2,
\]
which is exactly \eqref{eq:Gaussian_composition_rule_app}.
\end{proof}

\section{Lyapunov gauge covariance for the squeezed-reservoir example}
\label{app:Lyapunov_squeezed_stepwise}

Here we derive step by step the closed-form entries of the Lyapunov gauging covariance $S$ for the single-mode
drift matrix~\eqref{eq:A_DPA_squeezed}, starting from the Lyapunov equation
\begin{equation}
A S + S A^{\T} + D = 0,
\label{eq:Lyapunov_app}
\end{equation}
and we show how the EP-line formulas~\eqref{eq:S_EP_plus} follow by explicit substitution of~\eqref{eq:D_squeezed}.

\subsection{Explicit $C$, $C^\dagger C$, and its real/imaginary parts.}
\label{app:explicit_cc}

For a single jump operator $\hat L=c^{\T}\hat x$ the Ma--Woolley--Petersen matrix is simply
\begin{equation}
C=c^{\T}\in\mathbb C^{1\times 2}.
\end{equation}
From \eqref{eq:c_row} we have the explicit row vector
\begin{equation}
C
=
\frac{1}{\sqrt{2}}
\Big(\ \cosh r+e^{\ii\phi}\sinh r,\ \ \ii(\cosh r-e^{\ii\phi}\sinh r)\ \Big).
\label{eq:C_row_explicit}
\end{equation}
Write $C=(C_q,\ C_p)$, so
\begin{equation}
C_q=\frac{1}{\sqrt{2}}\big(\cosh r+e^{\ii\phi}\sinh r\big),\quad
C_p=\frac{\ii}{\sqrt{2}}\big(\cosh r-e^{\ii\phi}\sinh r\big).
\label{eq:C_components}
\end{equation}
Then
\begin{equation}
C^\dagger C
=
\begin{pmatrix}
|C_q|^2 & C_q^*\,C_p\\[0.3ex]
C_p^*\,C_q & |C_p|^2
\end{pmatrix}.
\label{eq:CdC_struct}
\end{equation}
The diagonal entries are
\begin{align}
|C_q|^2
&=\frac12\big(\cosh^2 r+\sinh^2 r+2\cosh r\sinh r\cos\phi\big)
=\frac12\big(\cosh(2r)+\sinh(2r)\cos\phi\big),
\label{eq:Cq_abs2}\\
|C_p|^2
&=\frac12\big(\cosh^2 r+\sinh^2 r-2\cosh r\sinh r\cos\phi\big)
=\frac12\big(\cosh(2r)-\sinh(2r)\cos\phi\big).
\label{eq:Cp_abs2}
\end{align}
For the off-diagonal entry, use \eqref{eq:C_components}:
\begin{align}
C_q^*C_p
&=
\frac{1}{2}\big(\cosh r+e^{-\ii\phi}\sinh r\big)\;
\ii\big(\cosh r-e^{\ii\phi}\sinh r\big)
\nonumber\\
&=\frac{\ii}{2}\Big(\cosh^2 r-\sinh^2 r-\cosh r\sinh r\,(e^{\ii\phi}-e^{-\ii\phi})\Big)
\nonumber\\
&=\frac{\ii}{2}\Big(1-2\ii\cosh r\sinh r\sin\phi\Big)
=\cosh r\sinh r\sin\phi+\frac{\ii}{2}
=\frac12\sinh(2r)\sin\phi+\frac{\ii}{2}.
\label{eq:CqCp}
\end{align}
Thus
\begin{equation}
C^\dagger C
=
\frac12
\begin{pmatrix}
\cosh(2r)+\sinh(2r)\cos\phi
&
\sinh(2r)\sin\phi+\ii
\\[0.6ex]
\sinh(2r)\sin\phi-\ii
&
\cosh(2r)-\sinh(2r)\cos\phi
\end{pmatrix}.
\label{eq:CdC_explicit}
\end{equation}
Taking real and imaginary parts entrywise gives
\begin{align}
\Ree[C^\dagger C]
&=
\frac{1}{2}
\begin{pmatrix}
\cosh(2r)+\sinh(2r)\cos\phi & \sinh(2r)\sin\phi\\
\sinh(2r)\sin\phi & \cosh(2r)-\sinh(2r)\cos\phi
\end{pmatrix},
\label{eq:Re_CdC_app}\\
\Imm[C^\dagger C]
&=
\frac{1}{2}
\begin{pmatrix}
0 & 1\\
-1 & 0
\end{pmatrix}
=
\frac{1}{2}\,\Sigma.
\label{eq:Im_CdC_app}
\end{align}
Including the overall rate $\kappa$ in \eqref{eq:ME_squeezed_res} amounts to replacing
\begin{equation}
C^\dagger C \ \mapsto\ \kappa\,C^\dagger C,
\quad\text{so that}\quad
\Ree[C^\dagger C]\mapsto \kappa\,\Ree[C^\dagger C],\ \ \Imm[C^\dagger C]\mapsto \kappa\,\Imm[C^\dagger C].
\end{equation}

\section{Derivation of the integral Lyapunov solution}
\label{app:Lyapunov_integral_derivation}

Assume $A$ is Hurwitz, i.e.\ $\Ree\,\lambda(A)<0$. Define
\begin{equation}
S:=\int_{0}^{\infty} e^{At}\,D\,e^{A^{\T}t}\,\dd t.
\label{eq:S_integral_app}
\end{equation}
The integral converges absolutely in operator norm because $e^{At}\to 0$ exponentially.

Differentiate the integrand:
\begin{equation}
\frac{\dd}{\dd t}\big(e^{At}D e^{A^{\T}t}\big)
=
A e^{At} D e^{A^{\T}t} + e^{At} D e^{A^{\T}t} A^{\T}.
\label{eq:integrand_derivative_app}
\end{equation}
Integrating \eqref{eq:integrand_derivative_app} from $0$ to $\infty$ and using dominated convergence yields
\begin{align}
A S + S A^{\T}
&=
\int_{0}^{\infty} \frac{\dd}{\dd t}\big(e^{At}D e^{A^{\T}t}\big)\,\dd t
=
\Big[e^{At}D e^{A^{\T}t}\Big]_{t=0}^{t=\infty}
=
-\,D,
\end{align}
since $e^{At}D e^{A^{\T}t}\to 0$ as $t\to\infty$ and the $t=0$ boundary gives $D$. Therefore
\begin{equation}
A S + S A^{\T} + D = 0,
\end{equation}
proving the integral representation \eqref{eq:S_integral_app}.

\paragraph{Uniqueness.}
If $A$ is Hurwitz and $AS+SA^{\T}=0$, then $F(t):=e^{-At}S e^{-A^{\T}t}$ satisfies $\dot F(t)=0$, hence is constant.
Taking $t\to\infty$ gives $F(t)\to 0$, so $S=0$. Thus the Lyapunov solution is unique.

\section{Vectorization solution of the one-mode Lyapunov equation}
\label{app:Lyapunov_vec_one_mode}

We show how the single-mode Lyapunov equation
\begin{equation}
A S + S A^{\T} + D = 0,
\label{eq:Lyapunov_app_one_mode}
\end{equation}
with $S=S^{\T}$ and $D=D^{\T}$, reduces to an explicit $3\times 3$ linear system by a vectorization strategy.

\subsection*{Vectorization and reduction to symmetric unknowns}

We use the standard identity
\begin{equation}
\vect(AXB)=(B^{\T}\!\otimes A)\,\vect(X),
\label{eq:vect_AXB_Lyap}
\end{equation}
and the column-stacking convention
\[
\vect\!\begin{pmatrix}x_{11}&x_{12}\\ x_{21}&x_{22}\end{pmatrix}
=
\begin{pmatrix}x_{11}\\ x_{21}\\ x_{12}\\ x_{22}\end{pmatrix}.
\]
Vectorizing \eqref{eq:Lyapunov_app_one_mode} gives
\begin{equation}
\vect(AS)+\vect(SA^{\T})+\vect(D)=0
\quad\Longleftrightarrow\quad
\bigl(I\otimes A + A\otimes I\bigr)\,\vect(S)=-\,\vect(D),
\label{eq:Lyapunov_vec_4x4_app}
\end{equation}
where we used $\vect(AS)=\vect(A S I)=(I\otimes A)\vect(S)$ and
$\vect(SA^{\T})=\vect(I S A^{\T})=(A\otimes I)\vect(S)$.

Since $S$ is symmetric, only three independent unknowns are needed. Define
\begin{equation}
s:=\begin{pmatrix}s_{11}\\ s_{12}\\ s_{22}\end{pmatrix},
\quad
y:=\begin{pmatrix}d_{11}\\ d_{12}\\ d_{22}\end{pmatrix},
\label{eq:s_y_defs_Lyap}
\end{equation}
and note that
\begin{equation}
\vect(S)=
\begin{pmatrix}s_{11}\\ s_{12}\\ s_{12}\\ s_{22}\end{pmatrix}
=:D_2\,s,
\quad
\vect(D)=
\begin{pmatrix}d_{11}\\ d_{12}\\ d_{12}\\ d_{22}\end{pmatrix}
=:D_2\,y,
\label{eq:duplication_Lyap}
\end{equation}
where
\begin{equation}
D_2=
\begin{pmatrix}
1&0&0\\
0&1&0\\
0&1&0\\
0&0&1
\end{pmatrix}.
\label{eq:D2_Lyap}
\end{equation}
To project the $4$-vector equation onto the symmetric subspace, introduce the linear map
\begin{equation}
L_2=
\begin{pmatrix}
1&0&0&0\\
0&\tfrac12&\tfrac12&0\\
0&0&0&1
\end{pmatrix},
\quad
L_2D_2=I_3,
\label{eq:L2_Lyap}
\end{equation}
which keeps the $(1,1)$ and $(2,2)$ entries and averages the duplicated off-diagonals.

Substituting \eqref{eq:duplication_Lyap} into \eqref{eq:Lyapunov_vec_4x4_app} and left-multiplying by $L_2$ yields the
equivalent $3\times 3$ system
\begin{equation}
M\,s=-\,y,
\quad
M:=L_2\bigl(I\otimes A + A\otimes I\bigr)D_2.
\label{eq:Lyapunov_3x3_reduction_app}
\end{equation}

\subsection*{Explicit $3\times 3$ matrix for $A\in\mathbb R^{2\times 2}$}

Write
\[
A=\begin{pmatrix}a&b\\ c&d\end{pmatrix}.
\]
A direct computation gives
\begin{equation}
I\otimes A + A\otimes I
=
\begin{pmatrix}
2a&b&b&0\\
c&a+d&0&b\\
c&0&a+d&b\\
0&c&c&2d
\end{pmatrix}.
\label{eq:Kronecker_sum_2x2_app}
\end{equation}
Applying $D_2$ merges the symmetric unknowns and produces
\begin{equation}
\bigl(I\otimes A + A\otimes I\bigr)D_2
=
\begin{pmatrix}
2a&2b&0\\
c&a+d&b\\
c&a+d&b\\
0&2c&2d
\end{pmatrix},
\label{eq:KD2_Lyap_app}
\end{equation}
and then applying $L_2$ averages the duplicated middle rows. Hence the reduced matrix $M$ in
\eqref{eq:Lyapunov_3x3_reduction_app} is
\begin{equation}
M=
\begin{pmatrix}
2a & 2b & 0\\
c & a+d & b\\
0 & 2c & 2d
\end{pmatrix}.
\label{eq:M_Lyapunov_3x3_app}
\end{equation}
Therefore, \eqref{eq:Lyapunov_app_one_mode} is equivalent (on symmetric unknowns) to
\begin{equation}
\begin{pmatrix}
2a & 2b & 0\\
c & a+d & b\\
0 & 2c & 2d
\end{pmatrix}
\begin{pmatrix}
s_{11}\\ s_{12}\\ s_{22}
\end{pmatrix}
=
-
\begin{pmatrix}
d_{11}\\ d_{12}\\ d_{22}
\end{pmatrix}.
\label{eq:Lyapunov_3x3_general_app}
\end{equation}
When $\det M\neq 0$ (equivalently, $\lambda_i(A)+\lambda_j(A)\neq 0$ for all eigenvalues), this system has a unique
solution, which can be obtained in closed form by standard elimination.

The three equations in \cref{eq:Lyapunov_3x3_general_app} are
\begin{align}
2a\,s_{11}+2b\,s_{12}&=-d_{11}, \label{eq:Lyap_eq1}\\
c\,s_{11}+(a+d)\,s_{12}+b\,s_{22}&=-d_{12}, \label{eq:Lyap_eq2}\\
2c\,s_{12}+2d\,s_{22}&=-d_{22}. \label{eq:Lyap_eq3}
\end{align}

Solve \eqref{eq:Lyap_eq1} and \eqref{eq:Lyap_eq3} for $s_{11}$ and $s_{22}$ in terms of $s_{12}$ (assuming $a\neq0$ and
$d\neq0$):
\begin{equation}
s_{11}=-\frac{d_{11}+2b\,s_{12}}{2a},
\quad
s_{22}=-\frac{d_{22}+2c\,s_{12}}{2d}.
\label{eq:s11_s22_in_terms_s12}
\end{equation}

Substitute \eqref{eq:s11_s22_in_terms_s12} into \eqref{eq:Lyap_eq2}:
\begin{align}
c\!\left(-\frac{d_{11}+2b\,s_{12}}{2a}\right)
+(a+d)\,s_{12}
+b\!\left(-\frac{d_{22}+2c\,s_{12}}{2d}\right)
&=-d_{12},
\nonumber\\
-\frac{c\,d_{11}}{2a}-\frac{bc}{a}s_{12}+(a+d)s_{12}-\frac{b\,d_{22}}{2d}-\frac{bc}{d}s_{12}
&=-d_{12}.
\label{eq:substituted_eq2}
\end{align}
Collect the $s_{12}$ terms:
\begin{equation}
\left((a+d)-bc\Big(\frac1a+\frac1d\Big)\right)s_{12}
=
-d_{12}+\frac{c\,d_{11}}{2a}+\frac{b\,d_{22}}{2d}.
\label{eq:collect_s12}
\end{equation}

Multiply \eqref{eq:collect_s12} by $2ad$ to clear denominators:
\begin{equation}
2ad\left((a+d)-bc\Big(\frac1a+\frac1d\Big)\right)s_{12}
=
-2ad\,d_{12}+cd\,d_{11}+ab\,d_{22}.
\label{eq:clear_denoms}
\end{equation}
Use $(\frac1a+\frac1d)=\frac{a+d}{ad}$ to simplify the prefactor:
\begin{align}
2ad\left((a+d)-bc\frac{a+d}{ad}\right)
&=2ad(a+d)-2bc(a+d)
\nonumber\\
&=2(a+d)(ad-bc)
=2(a+d)\det A.
\end{align}
Hence
\begin{equation}
s_{12}
=
\frac{ab\,d_{22}+cd\,d_{11}-2ad\,d_{12}}
{2(a+d)(ad-bc)}.
\label{eq:s12_closed_elimination}
\end{equation}

Finally, back-substitute \eqref{eq:s12_closed_elimination} into \eqref{eq:s11_s22_in_terms_s12}:
\begin{equation}
s_{11}=-\frac{d_{11}+2b\,s_{12}}{2a},
\quad
s_{22}=-\frac{d_{22}+2c\,s_{12}}{2d},
\label{eq:s11_s22_backsub}
\end{equation}
with $s_{12}$ given by \eqref{eq:s12_closed_elimination}. The resulting expressions extend by continuity to cases where
$a=0$ or $d=0$ provided the symmetric Lyapunov solution is unique (equivalently, the $3\times3$ matrix in
\eqref{eq:Lyapunov_3x3_general_app} is invertible).

\section{Single-mode complete positivity reduces to a determinant condition}
\label{app:single_mode_CP_det}

In this Appendix we show that, for one mode, the complete-positivity (CP)
constraint for a Gaussian map $(X_t,Y_t)$ collapses to a scalar determinant
inequality. We work with quadratures $\hat R=(\hat q,\hat p)^{\T}$ satisfying
\begin{equation}
[\hat R_j,\hat R_k]= i\,\Sigma_{jk},
\quad
\Sigma:=\begin{pmatrix}0&1\\-1&0\end{pmatrix},
\end{equation}
and with the standard CP condition for a Gaussian channel,
\begin{equation}
Y_t+\frac{i}{2}\Big(\Sigma- X_t\Sigma X_t^{\T}\Big)\succeq 0,
\label{eq:CP_condition_standard}
\end{equation}
where $X_t\in\mathbb R^{2\times 2}$ and $Y_t=Y_t^{\T}\in\mathbb R^{2\times 2}$.

\subsection{A $2\times 2$ identity: $X\Sigma X^{\T}=(\det X)\Sigma$}

Let $X=\begin{pmatrix}a&b\\ c&d\end{pmatrix}\in\mathbb R^{2\times2}$. A direct
multiplication yields
\begin{equation}
X\Sigma=
\begin{pmatrix}a&b\\ c&d\end{pmatrix}
\begin{pmatrix}0&1\\-1&0\end{pmatrix}
=
\begin{pmatrix}-b & a\\ -d & c\end{pmatrix},
\end{equation}
and therefore
\begin{align}
X\Sigma X^{\T}
&=
\begin{pmatrix}-b & a\\ -d & c\end{pmatrix}
\begin{pmatrix}a&c\\ b&d\end{pmatrix}
\nonumber\\
&=
\begin{pmatrix}
0 & ad-bc\\
-(ad-bc) & 0
\end{pmatrix}
=(\det X)\,\Sigma .
\label{eq:XSigmaXT_det}
\end{align}
This identity is special to $2\times 2$ matrices: $\Sigma$ represents the
area form on phase space, which is scaled by $\det X$.

\subsection{Reduction of the CP matrix to a one-parameter family}

Using \eqref{eq:XSigmaXT_det} in \eqref{eq:CP_condition_standard} gives
\begin{equation}
\Sigma- X_t\Sigma X_t^{\T}=\Sigma-(\det X_t)\Sigma=(1-\det X_t)\Sigma,
\end{equation}
so the CP matrix can be written as
\begin{equation}
Z_t:=Y_t+\frac{i}{2}\Big(\Sigma- X_t\Sigma X_t^{\T}\Big)
=Y_t+i\alpha_t \Sigma,
\quad
\alpha_t:=\frac{1-\det X_t}{2}.
\label{eq:Zt_def_alpha}
\end{equation}
Hence, in the single-mode case, the dependence of the quantum correction in
\eqref{eq:CP_condition_standard} on $X_t$ is entirely through the scalar
$\det X_t$.

\subsection{Positivity of $Z_t$ and the determinant inequality}

Write $Y_t$ explicitly as a real symmetric matrix,
\begin{equation}
Y_t=\begin{pmatrix}y_1&y_3\\ y_3&y_2\end{pmatrix},
\quad y_1,y_2,y_3\in\mathbb R,
\end{equation}
so that, using $i\alpha_t\Sigma=\begin{pmatrix}0&i\alpha_t\\ -i\alpha_t&0\end{pmatrix}$,
\begin{equation}
Z_t=
\begin{pmatrix}
y_1 & y_3+i\alpha_t\\
y_3-i\alpha_t & y_2
\end{pmatrix}.
\label{eq:Zt_explicit}
\end{equation}
Since $Z_t$ is Hermitian, $Z_t\succeq 0$ is equivalent to the nonnegativity of
its principal minors, i.e.\ $Z_{11}\ge 0$ and $\det Z_t\ge 0$ (and, for completeness,
also $Z_{22}\ge 0$, which follows automatically when $Z_{11}>0$ and $\det Z_t\ge 0$).
In our setting we additionally impose $Y_t\succeq 0$, which already enforces
$y_1\ge 0$ and $y_2\ge 0$. The genuinely new constraint from the quantum term is
therefore $\det Z_t\ge 0$. Computing the determinant of \eqref{eq:Zt_explicit} we find
\begin{align}
\det Z_t
&=y_1y_2-(y_3+i\alpha_t)(y_3-i\alpha_t)
\nonumber\\
&=y_1y_2-(y_3^2+\alpha_t^2)
=\det Y_t-\alpha_t^2.
\label{eq:detZt}
\end{align}
Thus $\det Z_t\ge 0$ is equivalent to
\begin{equation}
\det Y_t\ge \alpha_t^2=\Big(\frac{1-\det X_t}{2}\Big)^2.
\end{equation}
Collecting conditions, we obtain the single-mode reduction:
\begin{equation}
Y_t\succeq 0,
\quad
\det Y_t \ge \Big(\frac{1-\det X_t}{2}\Big)^2,
\label{eq:CP_single_mode_det_app}
\end{equation}
which matches Eq.~\eqref{eq:CP_single_mode_det} in the main text.

\begin{remark*}
Equation~\eqref{eq:CP_single_mode_det_app} shows that, for one mode, complete
positivity depends on $X_t$ only through the phase-space area scaling
$\det X_t$. In contrast, for $N>1$ modes the matrix constraint
$Y_t+\frac{i}{2}(\Sigma-X_t\Sigma X_t^{\T})\succeq 0$ does \emph{not} reduce to
a single scalar condition.
\end{remark*}

\section{Explicit one-mode Stein solution via vectorization}
\label{app:Stein_one_mode_vec}

We present a compact derivation of the closed-form solution of the one-mode Stein equation
\begin{equation}
S=X S X^{\T}+Y,
\label{eq:Stein_app_vec}
\end{equation}
using vectorization and a $3\times3$ reduction to the symmetric subspace.

\subsection{Vectorization and reduction to a $3\times3$ system}
\label{app:Stein_vec_3x3}

Write
\begin{equation}
X=\begin{pmatrix}a&b\\ c&d\end{pmatrix},\quad
Y=\begin{pmatrix}y_{11}&y_{12}\\ y_{12}&y_{22}\end{pmatrix},\quad
S=\begin{pmatrix}s_{11}&s_{12}\\ s_{12}&s_{22}\end{pmatrix}.
\label{eq:XYZS_vec}
\end{equation}
Vectorize \eqref{eq:Stein_app_vec}. Using the standard identity
\begin{equation}
\vect(AXB)=(B^{\T}\otimes A)\,\vect(X),
\label{eq:vect_AXB_vec}
\end{equation}
we obtain
\begin{equation}
\vect(S)=\vect(XSX^{\T})+\vect(Y)=(X\otimes X)\,\vect(S)+\vect(Y),
\label{eq:vec_Stein_step}
\end{equation}
hence
\begin{equation}
\bigl(I_4-X\otimes X\bigr)\,\vect(S)=\vect(Y).
\label{eq:vec_Stein_4x4}
\end{equation}

To restrict \eqref{eq:vec_Stein_4x4} to symmetric matrices, introduce the three-vectors of independent entries
\begin{equation}
s:=\begin{pmatrix}s_{11}\\ s_{12}\\ s_{22}\end{pmatrix},
\quad
y:=\begin{pmatrix}y_{11}\\ y_{12}\\ y_{22}\end{pmatrix}.
\label{eq:s_y_defs_vec}
\end{equation}
Then $\vect(S)$ and $\vect(Y)$ are obtained from $s$ and $y$ by the (fixed) duplication matrix
\begin{equation}
D_2=
\begin{pmatrix}
1&0&0\\
0&1&0\\
0&1&0\\
0&0&1
\end{pmatrix},
\quad
\vect(S)=D_2\,s,\quad \vect(Y)=D_2\,y.
\label{eq:D2_def_vec}
\end{equation}
Conversely, define the elimination matrix
\begin{equation}
L_2=
\begin{pmatrix}
1&0&0&0\\
0&\tfrac12&\tfrac12&0\\
0&0&0&1
\end{pmatrix},
\quad
L_2D_2=I_3,
\label{eq:L2_def_vec}
\end{equation}
so that $s=L_2\vect(S)$ and $y=L_2\vect(Y)$ whenever $S$ and $Y$ are symmetric.

Left-multiplying \eqref{eq:vec_Stein_4x4} by $L_2$ and substituting \eqref{eq:D2_def_vec} gives
\begin{equation}
\underbrace{L_2\bigl(I_4-X\otimes X\bigr)D_2}_{=:M}\,s=y.
\label{eq:svect_3x3_system_nosvect}
\end{equation}
A direct multiplication yields
\begin{equation}
M=
\begin{pmatrix}
1-a^2 & -2ab & -b^2\\
-ac & 1-(ad+bc) & -bd\\
-c^2 & -2cd & 1-d^2
\end{pmatrix},
\label{eq:M_explicit_vec}
\end{equation}
so the Stein equation on symmetric matrices reduces to the $3\times3$ linear system
\begin{equation}
M
\begin{pmatrix}s_{11}\\ s_{12}\\ s_{22}\end{pmatrix}
=
\begin{pmatrix}y_{11}\\ y_{12}\\ y_{22}\end{pmatrix}.
\label{eq:Stein_3x3_vec_final}
\end{equation}

\subsection{Closed form by inversion of the $3\times3$ matrix}
\label{app:Stein_vec_closed}

Define the drift invariants
\begin{equation}
\tau:=\Tr X=a+d,\quad
\Delta:=\det X=ad-bc,
\label{eq:tauDelta_vec}
\end{equation}
and the denominator
\begin{equation}
\mathcal D:=(1-\Delta)\,(1-\tau+\Delta)\,(1+\tau+\Delta).
\label{eq:D_denom_vec}
\end{equation}
One finds $\det M=\mathcal D$. For $\mathcal D\neq0$, the solution is unique and given by
\begin{equation}
\begin{pmatrix}s_{11}\\ s_{12}\\ s_{22}\end{pmatrix}
=
\frac{1}{\mathcal D}\,\adj(M)
\begin{pmatrix}y_{11}\\ y_{12}\\ y_{22}\end{pmatrix},
\end{equation}
since ${M^{-1}= \frac{1}{\mathcal D} \adj(M)}$.
Carrying out the adjugate multiplication yields
\begin{align}
s_{11}
&=\frac{1}{\mathcal D}\Big[
\big(ad^{3}-ad-bcd^{2}-bc-d^{2}+1\big)\,y_{11}
+\big(-2abd^{2}+2ab+2b^{2}cd\big)\,y_{12}
+\big(ab^{2}d-b^{3}c+b^{2}\big)\,y_{22}
\Big],
\\[0.5ex]
s_{12}
&=\frac{1}{\mathcal D}\Big[
\big(-acd^{2}+ac+bc^{2}d\big)\,y_{11}
+\big(a^{2}d^{2}-a^{2}-b^{2}c^{2}-d^{2}+1\big)\,y_{12}
+\big(-a^{2}bd+ab^{2}c+bd\big)\,y_{22}
\Big],
\\[0.5ex]
s_{22}
&=\frac{1}{\mathcal D}\Big[
\big(ac^{2}d-bc^{3}+c^{2}\big)\,y_{11}
+\big(-2a^{2}cd+2abc^{2}+2cd\big)\,y_{12}
+\big(a^{3}d-a^{2}bc-a^{2}-ad-bc+1\big)\,y_{22}
\Big].
\end{align}

\subsection{Stability $\spr(X)<1$ in entry form and positivity of $\mathcal D$}
\label{app:Stein_vec_stability}

For $2\times2$ drift, $\spr(X)<1$ is equivalent by the Jury (Schur) criterion to the inequalities for the characteristic polynomial
$\lambda^2-\tau\lambda+\Delta$~\cite{DuHsiauLiMalkin2007}:
\begin{equation}
\spr(X)<1
\quad\Longleftrightarrow\quad
1-\Delta>0,\quad
1-\tau+\Delta>0,\quad
1+\tau+\Delta>0,
\end{equation}
which imply $\mathcal D>0$.

\end{document}